  \documentclass[prc,showpacs,eqsecnum,floatfix,multicol,nofootinbib]{revtex4-2}

 \usepackage{axodraw2}   
 \usepackage{pstcol}   
 \usepackage{amsmath}  
 \usepackage{amsfonts}
 \usepackage{amssymb}  
 \usepackage{graphicx}  
 \usepackage{pstricks}  
 \usepackage{bm}    
\DeclareMathOperator{\Tr}{Tr}
\begin{document}
 
%\title{ 
%Diquark Feynman Propagator in Mixed Nuclear-Quark Matter \\
% Nucleon-Quark Diquark-exchange Interaction }   
\title{ Nucleon-Quark Diquark-exchange Interaction
 \footnote{For earlier version see: "Diquark Feynman Propagator in Mixed Nuclear-Quark 
 Matter; Nucleon-Quark Diquark-exchange Interaction", Report THEF 25-01, 
 {\it https://nn-online.org/eprints (2025).}}
 }                   
\author{Th.A.\ Rijken}
\affiliation{ Institute of Mathematics, Astrophysics, and Particle Physics \\
 Radboud University, Nijmegen, The Netherlands }               

\date{version of: \today}

\begin{abstract}                                       
In this note the nucleon-quark diquark-exchange interaction is derived
using the Feynman-propagator for axial-vector diquark (D) exchange.
The Feynman diquark-propagator is derived and the result 
is a (-)-sign difference w.r.t. quark-antiquark exchange.          
This is due to the (-)-sign for a closed fermion loop,
present for {\it e.g} vector and axial-vector quark-antiquark exchange,       
but absent in the case of D-exchange.
The calculations in these notes follow closely those for vacuum polarization
in the literature. 
Taking into account that the diquark D is a color $\{\bar{3}\}_c$-state 
gives a factor +2. The result is an effective axial-vector 
diquark propagator in momentum space
\begin{eqnarray*}
i(\widetilde{\Delta})^{ab}_{\mu\nu}(k) &=& 
+2i\delta_{ab}\frac{\left(\eta_{\mu\nu}-k_\mu k_\nu/m_D^2\right)} 
{k^2-m_D^2+i\epsilon}.
\end{eqnarray*}
Application to $QN \rightarrow NQ$ gives a repulsive potential for the
axial-vector $\gamma_5\gamma_\mu$ NDQ-coupling, which has been used in
mixed nuclear-quark matter calculations \cite{YYR24}.
 \end{abstract}
\pacs{13.75.Cs, 12.39.Pn, 21.30.+y}

 \maketitle

%---------------------------------------------------------------------------------
%\twocolumngrid                                          
%------------------------------------------------------------------------------
%------------------------------------------------------------------------------
%\begin{flushleft}
%\rule{16cm}{0.5mm}
%\end{flushleft}
%------------------------------------------------------------------------------
\section{Introduction} \label{intro.diqrk}     
In these notes we derive the diquark (D) propagator in the context of the standard
quantum field theory (QFT). The axial-vector diquark field originates from presentation
of the proton current $\eta^{(1)}(x)$ in terms of three quarks 
\cite{Ioffe81,Ioffe83}. It is pointed out that there is an important difference
between quark-quark objects and mesons, which are quark-antiquark states in the quark-model. 
This shows up in the Wick-theorem giving a sign difference. Application to the
process $ N + Q \rightarrow Q +N$ for diquark-exchange leads to a (universal) axial-vector
repulsive interaction.

\noindent The content of these notes is as follows.
In section~\ref{qft.diqrk} the diquark-field and the diquark propagator are defined.
The latter using the standard formalism in field theory \cite{BD65,Mandl61,Schweber64,NO90}.
In section~\ref{app:propa} the diquark propagator is given using its composite quark-quark structure,
exploiting the Wick-theorem. Here, we note the important difference with for example 
$\rho$ and $A_1$ exchange within the quark-model.
In section~\ref{app:disp.a} the calculation in the momentum space is
carried out, using the Pauli-Villars regularization method \cite{P-V.1949}. We follow the treatment 
in \cite{BD65} for the vacuum polarization and renormalize the diquark coupling.
This results in a dispersion presentation of the diquark propagator.
In section~\ref{sec:diquarkpot} for $NQ \rightarrow QN$ the diquark exchange is worked out. 
In section~VI we give a short discussion and conclusion.

\noindent In Appendix~\ref{app:FD-WE} the Wick-expansion in relation with the Feynman
propagator is reviewed in some detail.
In Appendix~\ref{app:scalar-exchange} details for meson-exchange
are given, showing the important sign difference with di-quark exchange.
%Appendix~\ref{app:misc} contains an alternative calculation of the Dirac matrix elements,
%with the same result.
 In Appendix~\ref{app:scal} the potential for a scalar-diquark and pseudoscalar-diquark exchange 
 are derived. They arise from a second proton three-quark current $\eta^{(2)}(x)$
 discussed in \cite{Ioffe83}. 
%In Appendix~\ref{app:DQEXCH-PF} the diquark-exchange contact interaction is introduced 
%in the framework of the partition functional, with a sign consistent with the Feynman
%propagator prescription.
%------------------------------------------------------------------------------
%  not in this version in nn-online! 
%------------------------------------------------------------------------------
 In Appendix~\ref{app:DQEXCH-PF} the diquark-exchange contact NQ-interaction is considered in the 
 setting of the thermodynamic grand partition functional for mixed nuclear-quark matter, 
 with a sign consistent with the Feynman propagator prescription.

%------------------------------------------------------------------------------
\section{QFT Second-quantization Formalism} \label{qft.diqrk}     
%------------------------------------------------------------------------------
The triquark representation of the nucleon current in the context of the QCD sum rules has been 
introduced in \cite{Ioffe81} 
\begin{equation}
	\eta^{(1)}_N(x) = \left[\widetilde{q}^a(x) C \gamma^\mu q^b(x)\right] 
	\gamma_5\gamma_\mu q^c(x) \epsilon^{abc},
\label{app:propaa}\end{equation}
where C is the charge-conjugation matrix in Dirac-spinor space, and a,b,c denote the color
indices.
The composite diquark (D) field $\chi^a_\mu(x)$ is introduced in \cite{YYR24} using the current by 
writing
\begin{eqnarray}
	\eta^{(1)}_N(x) &=& (\hbar c)^2 \gamma_5\gamma^\mu q^a(x)\cdot \chi^a_\mu(x),\ \
	\chi^a_\mu(x) \equiv \epsilon^{abc} \widetilde{q}^b(x) C \gamma_\mu q^c(x)/(\hbar c)^2.
\label{app:propab}\end{eqnarray}

 The diquark D-field Feynman propagator is 
 \begin{eqnarray}
	 i(\Delta_F)_{\mu\nu}^{ab}(x'-x) &=& 
	 (0|T\left[\chi^a_\mu(x') \chi^{b \dagger}_\nu(x)\right]|0)
 \label{app:propa1}\end{eqnarray}
where the diquark fields are
\footnote{ Here and in most of the following we use units $\hbar=1, c=1$. }
\begin{eqnarray}
	\chi_\mu^a(x) &=& \widetilde{q}^c(x) C\gamma_\mu q^d(x)\ \varepsilon^{acd}\ ,\ 
	\chi_\mu^{b \dagger}(x) = -\bar{q}^e(x) \gamma_\mu C\ \widetilde{\bar{q}}^f(x)\ \varepsilon^{bef}
 \label{app:propa2}\end{eqnarray}
 The Dirac indices are contracted, and C is the charge conjugation Dirac matrix which satisfies 
 $C^{-1}\gamma_\mu C=-\gamma^T_\mu$ \cite{BD65}. 
 Note that because $\gamma_0 C \gamma_0 = -C$ the diquark field is an axial-vector field.
 The Dirac equation reads $\gamma^\mu\partial_\mu q(x)= -im q(x)$ and for the transposed
 $\partial^\mu\widetilde{q}(x)\widetilde{\gamma}_\mu= -im\widetilde{q}(x)$.
 \begin{eqnarray}
	 \partial^\mu \chi^a_\mu(x) &=& \left[\left(\partial^\mu\widetilde{q}^c(x)\right) C\gamma_\mu q^d(x)
	 +\widetilde{q}^c C\gamma_\mu \partial^\mu q^d(x)\right]\ \varepsilon^{acd} =0,
 \label{app:propa7}\end{eqnarray}
 as follows from $\partial^\mu\widetilde{q}(x)C\gamma_\mu = +im\ \widetilde{q}(x)C$,
due to  the properties of the charge conjugation matrix $C$ \cite{BD65}. It follows that
\begin{eqnarray}
	\partial^\mu(\Delta_F)^{ab}_{\mu\nu}(x-x') &=&
(0|\left[\chi^a_0(x),\chi_\nu^{b \dagger}(x')\right]\bigr|_{x_0=x'_0}|0)=0
 \label{app:propa8}\end{eqnarray}
since this is $\propto \Delta(x-x')\bigr|_{x_0=x'_0}=0$.

 \noindent Suppressing the color indices, the plane wave expansion of the quark field, 
 see {\it e.g.} \cite{BD65}, reads
 \begin{eqnarray}
	 q(x) &=& \sum_s \int\frac{d^3p}{(2\pi)^{3/2}} \sqrt{\frac{m}{E_p}} \left[
		 b(p,s)\ u(p,s)\ e^{-i p.x} + d^\dagger(p,s)\ v(p,s)\ e^{i p.x}\right], \\
	 \bar{q}(x) &=& \sum_s \int\frac{d^3p}{(2\pi)^{3/2}} \sqrt{\frac{m}{E_p}} \left[
		 b^\dagger(p,s)\ \bar{u}(p,s)\ e^{i p.x} + d(p,s)\ \bar{v}(p,s)\ e^{-i p.x}\right], \\
 \label{app:propa3a}\end{eqnarray}
 It is convenient to introduce the positive and negative frequency parts
 \begin{eqnarray}
	 q^{(+)}(x) &=& \sum_s \int\frac{d^3p}{(2\pi)^{3/2}} \sqrt{\frac{m}{E_p}} \left[
		 b(p,s)\ u(p,s)\ e^{-i p.x} \right], \\
	 q^{(-)}(x) &=& \sum_s \int\frac{d^3p}{(2\pi)^{3/2}} \sqrt{\frac{m}{E_p}} \left[
		 d^\dagger(p,s)\ v(p,s)\ e^{i p.x}\right], 
 \label{app:propa3b}\end{eqnarray}
 which are the annihilation and creation operators for a quark and a anti-quark respectively.
 Similarly
 \begin{eqnarray}
	 \bar{q}^{(-)}(x) &=& \sum_s \int\frac{d^3p}{(2\pi)^{3/2}} \sqrt{\frac{m}{E_p}} \left[
		 b^\dagger(p,s)\ \bar{u}(p,s)\ e^{i p.x} \right], \\
	 \bar{q}^{(+)}(x) &=& \sum_s \int\frac{d^3p}{(2\pi)^{3/2}} \sqrt{\frac{m}{E_p}} \left[
		 d(p,s)\ \bar{v}(p,s)\ e^{-i p.x}\right],
 \label{app:propa3c}\end{eqnarray}
 which are the creation and annihilation operators for a quark and a anti-quark respectively.
 The vacuum is defined as
 \begin{equation}
	 q^{(+)}(x)|0) = \bar{q}^{(+)}(x) |0) =0.                                     
 \label{app:propa3d}\end{equation}
 The annihilation and creation operators $b(p,s)$ and $d^\dagger(p,s)$ satisfy
 the anti-commutation relations unequal to zero are
 \begin{subequations}\label{app:propa4}
 \begin{eqnarray}
 \left\{b(p,s), b^\dagger(p',s')\right\} &=& \delta_{ss'} \delta^3({\bf p}-{\bf p}'), \\
 \left\{d(p,s), d^\dagger(p',s')\right\} &=& \delta_{ss'} \delta^3({\bf p}-{\bf p}'),    
 \end{eqnarray}\end{subequations}
 The anti-commutator is \cite{BD65}
 \begin{subequations}\label{app:propa5a}   
 \begin{eqnarray}
% && \bigl\{q^a_\alpha(x), q^{b \dagger}_\beta(x')\bigr\} = -i \delta_{a,b} 
% \left[S(x-x')\ \gamma^0\right]_{\alpha\beta}, \\
	 && \bigl\{q^a_\alpha(x), \bar{q}^b_\beta(x')\bigr\} = -i \delta_{a,b} 
	 S(x-x')_{\alpha\beta}, \\
	 && S_{\alpha\beta}(z) = (i\gamma^\mu\partial_\mu + m_Q)\ \Delta(z,m_Q^2).
 \end{eqnarray}\end{subequations}
 and the vacuum expectation of T-product is related to the quark Feynman-propagator, 
	 see \cite{BD65} section 13.6,
 \begin{subequations}\label{app:propa5b}   
 \begin{eqnarray}
% && (0|T\left[q^a_\alpha(x) q^{b \dagger}_\beta(x')\right]|0) = 
% -i \left[S_F(x-x')\ \gamma^0\right]_{\alpha\beta}, \\
 && (0|T\left[q^a_\alpha(x) \bar{q}^b_\beta(x')\right]|0) = 
	 -i \delta_{ab}S_F(x-x')_{\alpha\beta}, \\
 && S_{F \alpha\beta}(z) = (i\gamma^\mu\partial_\mu + m_Q)\ \Delta_F(z,m_Q^2).
 \end{eqnarray}\end{subequations}
	 For the invariant function $S_F(x-x')$ see \cite{Schweber64,BD65}. 
%For notational convenience we use in the following the abrieviation:
% \begin{equation}
% S_F^0(x-x') \equiv \left[S_F(x-x')\ \gamma^0\right]. 
% \label{app:SF0}\end{equation}
% We note that
% \begin{eqnarray}
%	 && (0|T\left[q^a_\alpha(x) q^{b \dagger}_\beta(x')\right]|0) = 
%	  (0|T\left[q^a_\alpha(x) \bar{q}^b_{\beta'}(x')\right]|0)\ (\gamma^0)_{\beta',\beta}  
% \label{app:propa55b}\end{eqnarray}
 The hermitian conjugate of 
$\chi_\mu^a(x) = \widetilde{q}^c(x) C\gamma_\mu q^d(x)\ \varepsilon^{acd}$   is 
 \begin{subequations}\label{intro:misc.1}
\begin{eqnarray}
%\chi_\mu^a(x) &=& \widetilde{q}^c(x) C\gamma_\mu q^d(x)\ \varepsilon^{acd}, \\ 
	\chi_\mu^{a \dagger}(x) &=& \widetilde{q}^{d\dagger}(x)\left(C\gamma_\mu\right)^\dagger 
q^{c \dagger}(x)\ \varepsilon^{acd} =
\widetilde{\bar{q}}^d(x) \gamma_\mu C\ \bar{q}^c\ \varepsilon^{acd},
\end{eqnarray}\end{subequations}
which will be used extensively in the rest of these notes.

%--------------------------------------------------------------------------
 %For the $S(x-x')$ function we use the sign
 %of \cite{NO90} which has a (-)-sign w.r.t. \cite{Schweber64}
%Using the notation $d^3\tilde{p} \equiv [d^3p/(2\pi)^{3/2}]\ \sqrt{m/E_p}$, 
%--------------------------------------------------------------------------
 \section{Diquark Propagator} \label{app:propa}
%--------------------------------------------------------------------------
% \begin{center}
%	 \fbox{ \begin{minipage}[b]{16cm}
% \vspace{2mm}
%%--------------------------------------------------------------------------
% \begin{center}
%	 {\blue \underline{Application Wick-theorem}:}
% \end{center}
% \vspace{-5mm}
%%--------------------------------------------------------------------------
 From the detailed derivation of the Wick-expansion in Appendix~\ref{app:FD-WE}
 for the axial-vector diquark propagator 
 we have 
\begin{eqnarray}
	\left(0|T\left[\chi^a_\mu(x') \chi^{b \dagger}_\nu(x)\right]|0\right) \Rightarrow
	&& X_{\mu\nu}= -\varepsilon^{acd}\varepsilon^{bef} \bigg[
		(0|T\left[q^d_\beta(x')\bar{q}^e_\kappa(x)\right]|0) 
		(0|T\bigl[q^c_\alpha(x')\bar{q}^f_\lambda(x)\bigr]|0) 
		\nonumber\\ &&
	       -(0|T\left[q^c_\alpha(x')\bar{q}^e_\kappa(x)\right]|0) 
	       (0|T\bigl[q^d_\beta(x')\bar{q}^f_\lambda(x)\bigr]|0)\biggr]
	O^{\alpha\beta}_\mu O^{\prime\kappa\lambda}_\nu
\label{app:propa3e}\end{eqnarray}
 \vspace{-5mm}

with 
 \begin{eqnarray*}
 O^{\alpha\beta}_\mu = (C\gamma_\mu)^{\alpha\beta}\ ,\ 
 O^{\prime\kappa\lambda}_\nu = (\gamma_\nu C)^{\kappa\lambda}
\end{eqnarray*}
 The color factors in (\ref{app:propa3e}) are
\begin{eqnarray}
	&& -\varepsilon^{acd}\varepsilon^{bef}\delta_{de}\delta_{cf}=+2\delta_{ab},\ \ {\rm and}\ \
  -\varepsilon^{acd}\varepsilon^{bef}\delta_{ce}\delta_{df}=-2\delta_{ab}.  
\label{app:propa5f}\end{eqnarray}
 \vspace{-5mm}

which give 

 \vspace{-5mm}
\begin{eqnarray}
%\left(0|T\left[\chi^a_\mu(x') \chi^{b \dagger}_\nu(x)\right]|0\right) =           
	&& X_{\mu\nu}= +2\delta_{ab} \bigg[
		(0|T\left[q_\beta(x')\bar{q}_\kappa(x)\right]|0) 
		(0|T\bigl[q_\alpha(x')\bar{q}_\lambda(x)\bigr]|0) 
		\nonumber\\ &&
	       +(0|T\left[q_\alpha(x')\bar{q}_\kappa(x)\right]|0) 
	       (0|T\bigl[q_\beta(x')\bar{q}_\lambda(x)\bigr]|0)\biggr]
	O^{\alpha\beta}_\mu O^{\prime\kappa\lambda}_\nu
\label{app:propa4e}\end{eqnarray}
 where $\left(0| .... |0\right)$ is diagonal in color, and a single component has to be used.
% For details on the Wick-expansion in (\ref{app:propa3e}), see Appendix~\ref{app:FD-WE}. \\
Using the Feynman propagators \cite{BD65} we obtain
 \begin{subequations}\label{app:propa5e}
\begin{eqnarray}
	&&  (0|T\left[q^d_\beta(x')\bar{q}^e_\kappa(x)\right]|0) 
	(0|T\bigl[q^c_\alpha(x')\bar{q}^f_\lambda(x)\bigr]|0) = 
	-\delta_{de}\delta_{cf} S_{F\beta\kappa}(x'-x)\ S_{F\alpha\lambda}(x'-x), 
	\label{app:prop5e.a}\\ &&
	       (0|T\left[q^c_\alpha(x')\bar{q}^e_\kappa(x)\right]|0) 
		(0|T\bigl[q^d_\beta(x')\bar{q}^f_\lambda(x)\bigr]|0) =
		-\delta_{ce}\delta_{df} S_{F\alpha\kappa}(x'-x)\ S_{F\beta\lambda}(x'-x)     
\end{eqnarray}\end{subequations}
 The result for $X_{\mu\nu}$ is
 \footnote{{\bf Note: the factor 2 due to color!}}$^)$
\begin{eqnarray}
 X_{\mu\nu} &=& -2\delta_{ab}\biggl[
S_{F\beta\kappa}(x'-x)\ S_{F\alpha\lambda}(x'-x)+ 
S_{F \alpha\kappa}(x'-x)\ S_{F\beta\lambda}(x'-x) \biggr]\ 
(C\gamma_\mu)_{\alpha\beta} (\gamma_\nu C)_{\kappa\lambda}
	\nonumber\\ &=& -2\delta_{ab}\
	Tr\left[\gamma_\mu S_F \gamma_\nu (C\widetilde{S}_FC)+S_F\gamma_\nu (C\widetilde{S}_FC)\ \gamma_\mu\right]
	=-4\delta_{ab} Tr\left[\gamma_\mu S_F \gamma_\nu (C\widetilde{S}_FC)\right]
 \label{app:propa5g}\end{eqnarray}
where in last lines we used the shorthand notation 
 $S_{F\alpha\beta} \equiv S_{F\alpha\beta}(x'-x)$.  Now,
 \begin{eqnarray*}
	 && C\widetilde{S}_F C = C\left(i\widetilde{\gamma}\cdot\partial +m_Q\right)\ C =
	 -\left(i\gamma\cdot\partial -m_Q\right), 
 \end{eqnarray*}
 \vspace{-5mm}

 giving      
\begin{eqnarray}
 X_{\mu\nu} &=& 
4\delta_{ab} Tr\left[\gamma_\mu \left(i\gamma\cdot\partial^y+m_Q\right) \gamma_\nu
\left(i\gamma\cdot\partial^z-m_Q\right)\right]\ \Delta_F(y)\cdot\Delta_F(z)   
\nonumber\\ &=& 
4\delta_{ab} \Tr\left[-\gamma_\mu\gamma_\alpha\gamma_\nu\gamma_\beta \partial^y_\alpha\partial^z_\beta
-\gamma_\mu\gamma_\nu m_Q^2 \right]\ \Delta_F(y)\cdot\Delta_F(z)    
 \label{app:propa5h}\end{eqnarray}
 \vspace{-5mm}
 Using 
 \begin{eqnarray*}
 && \Tr\bigl[\gamma_\mu\gamma_\alpha\gamma_\nu \gamma_\beta\bigr]= 
  4\left( \eta_{\mu\alpha}\eta_{\nu\beta} 
 -\eta_{\mu\nu}\eta_{\alpha\beta} +\eta_{\mu\beta}\eta_{\nu\alpha}\right)
 \end{eqnarray*}
 \vspace{-5mm}

 we obtain
\begin{eqnarray}
 X_{\mu\nu} &=& 
	-16\delta_{ab} \left[\left(\partial^y_\mu\partial^z_\nu+\partial^y_\nu\partial^z_\mu\right)
	-\partial^y_\alpha \partial_z^\alpha \eta_{\mu\nu}
	+\eta_{\mu\nu} m_Q^2 \right]\ \Delta_F(y)\cdot\Delta_F(z).
 \label{app:propa5i}\end{eqnarray}
 We introduced temporarily the differentiation variables $y,z$, which in the end will be put to $y=z=x'-x$.\\

 {\bf Remark: For $q\bar{q}$-exchange the Wick-theorem gives a (-)-sign similar to that for a closed
 fermion-loop.}\\

\noindent  The diquark field Feynman propagator is 
 \footnote{We follow the conventions of \cite{BD65}, Appendix B and C in part II, 
 in definition functions $\Delta_F(x)$ etc and Feynman rules.}
 \begin{eqnarray}
 i(\Delta_F)_{\mu\nu}^{ab}(x'-x) &=& 
	 (0|T\left[\chi^a_\mu(x') \chi^{b \dagger}_\nu(x)\right]|0) =
	 -16\delta_{ab} \biggl\{\biggl[\left(\partial_\mu\Delta_F(y)\cdot \partial_\nu\Delta_F(z) 
	 +\partial_\nu\Delta_F(y)\cdot\partial_\mu\Delta_F(z)\right)
	 \nonumber\\ && 
	 -\partial_\alpha\Delta_F(y)\cdot \partial^\alpha \Delta_F(z)\eta_{\mu\nu}
	 +\eta_{\mu\nu} m_Q^2 \Delta_F(y)\cdot\Delta_F(z)\biggr] \biggr\}_{y=z=x'-x}
 \label{app:propa6a}\end{eqnarray}
%--------------------------------------------------------------------------
% \end{minipage} }\\
% \end{center}
%--------------------------------------------------------------------------
%%------------------------------------------------------------------------------
%% figuur 1
%\begin{center}
% \begin{figure}[hbt]
%\centering
%%\resizebox{7.25cm}{5.75cm}
%%\resizebox{7.25cm}{6.75cm}
% \resizebox{9.25cm}{6.75cm}
%%{\includegraphics[0in,0in][8in,10in]{Fig.diquark-exchange.IV.ps}}
%%{\includegraphics*[width=16cm,height=8cm]{Fig.diquark-exchange.IV.ps}}
%%{\includegraphics[180,375][400,555]{Fig.diquark-exchange.IV.ps}}
% {\includegraphics[140,375][360,555]{Fig.diquark-exchange.IV.ps}}
% \vspace{+1.0cm}
%\caption{Diquark-exchange NQ $\rightarrow$ QN. 
%Panel (a): $\chi^a_\mu$-exchange. Panel (b): D-exchange.}
%\label{fig:diqrk-exchange.IV}
%\end{figure}
%\end{center}
%%---------------------------------------------------------------------------------
%%--------------------------------------------------------------------------
% axodraw in Tex:
%--------------------------------------------------------------------------
 \begin{figure}[hbtp!]
%--------------------------------------------------------------------------
 % (a) axial-vector $\chi^a_\mu$-exchange diagram:	 
%--------------------------------------------------------------------------
 \begin{center} \begin{picture}(175,300)(75,0)
%\begin{center} \begin{picture}(200,175)(0,75)
%\begin{center} \begin{picture}(200,175)(0,15)
 \SetPFont{Helvetica}{9}
 \SetScale{1.0} \SetWidth{1.5}
 \SetOffset(0,0)
 \SetOffset(-110,0)
 \SetOffset(-70 ,0)
 \SetOffset(-30 ,0)
 \ArrowLine(150,15)(150,75)   
 \ArrowLine(150,75)(150,135)   
 \Vertex(150,75){3}
 \Vertex(90, 75){3}
%\DashLine(75,90)(75,150){2}
%\Photon(75,150)(75,122){1}{3}
%\Text( 75,120)[]{$\bigotimes$}
%\Photon(75,90)(75,118){1}{3}
 \ArrowLine(90,15)(90,75)   
 \ArrowLine(90,75)(90,135)   
%---------------------------------------------------------------------------
 \SetWidth{0.3}
 \SetWidth{1.0}
 \SetColor{Blue}
 \Photon( 90,75)(150,75){1}{6}
%\ArrowArc(120,25 )(60,60,120)  
%\ArrowArcn(120,125)(60,-60,-120)
 \SetColor{Black}
 \SetWidth{1.5}
%\DashArrowArc(75,120)(30,-90,90){1}
%\DashArrowArcn(75,120)(30,-90,90){1}
%---------------------------------------------------------------------------
 \Text(150,0  )[]{$N,p_1$}
 \Text(150,155)[]{$Q,p_1'$}
 \Text(90,  0)[]{$Q,p_2$}
 \Text(90,155)[]{$N,p_2'$}
%\Text(120,55)[]{$p$}
%\Text(120,95)[]{$q$}
	 \Text(185,75)[]{$=$}
 \Text(120,-5)[]{(a)}
%--------------------------------------------------------------------------
% (b) Diquark-exchange diagram:
%---------------------------------------------------------------------------
  \SetOffset(140,0)
  \SetOffset(100,0)
  \SetOffset(140,0)
% \SetOffset(0,-110)
  \ArrowLine(120,15)(120,75)   
  \ArrowLine(120,75)(120,135)   
% \Vertex(180,75){3}
  \Vertex(60, 75){3}
  \Vertex(120,75){3}
% \Vertex(90, 75){3}
 \SetColor{Red}
%\Photon(180,75)(150,75){1}{3}
%\Photon(60,75)(90,75){1}{3}
  \SetColor{Black}
  \ArrowLine(60,15)(60,75)   
  \ArrowLine(60,75)(60,135)   
%---------------------------------------------------------------------------
 \SetWidth{0.5}
 \SetWidth{1.0}
%\SetColor{Red}
%\CArc(75,120)(30,-90,90)
 \SetColor{Red}  
%CArc( 25,120)(60,-30,30)  
 \ArrowArc( 90,25 )(60,60,120)  
%\CArc(125,120)(60,150,210)
 \ArrowArcn( 90,125)(60,-60,-120)
 \SetColor{Black}
 \SetWidth{1.5}
%\DashArrowArc(75,120)(30,-90,90){1}
%\DashArrowArcn(75,120)(30,-90,90){1}
%---------------------------------------------------------------------------
 \SetWidth{1.5}
  \Text(120,0  )[]{$N,p_1$}
  \Text(120,155)[]{$N,p_1'$}
  \Text(60,  0)[]{$Q,p_2$}
  \Text(60,155)[]{$Q,p_2'$}
	 \Text( 90,50)[]{$p$}
 \Text( 90,100)[]{$q$}
 \Text( 90,-5)[]{(b)}
%---------------------------------------------------------------------------
 \end{picture} \end{center}
% \vspace{-5mm}
 \vspace{+0.5cm}
	\caption{Diquark-exchange NQ $\rightarrow$ QN. 
	Panel (a): $\chi^a_\mu$-exchange. Panel (b): QQ-pair (D) exchange.}
	\label{fig:diqrk-exchange.IV}
\end{figure}
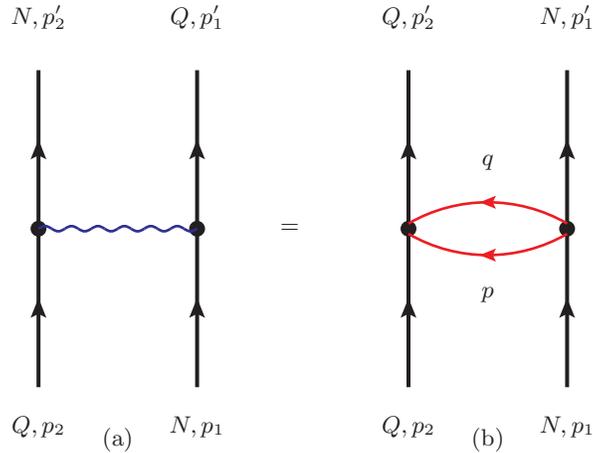
%---------------------------------------------------------------------------------

 %-----------------------------------------------------------------------------------------
 \section{Propagator Dispersion relation } \label{app:disp.a}
 %-----------------------------------------------------------------------------------------
 The spectral representation of the Diquark Feynman propagator is 
 \begin{eqnarray}
	 i(\Delta_F)_{\mu\nu}^{ab}(x'-x) &=& 
	 (0|T\left[\chi^a_\mu(x') \chi^{b \dagger}_\nu(x)\right]|0)
	 = i\int_{s_0}^\infty ds\ (\Delta_F)^{ab}_{\mu\nu}(x'-x;s)\ \rho(s).
 \label{app:disp1}\end{eqnarray}
 %-----------------------------------------------------------------------------------------
%We consider the (basic) integral
%\begin{eqnarray}
%\widetilde{I}_0(k) &=& \int\frac{d^4p}{(2\pi)^4} \int\frac{d^4q}{(2\pi)^4} 
%(2\pi)^4\delta^4(p+q-k) 
%\cdot\nonumber\\ && \times
%\left[p^2-m_Q^2+i\epsilon\right]^{-1} \left[q^2-m_Q^2+i\epsilon\right]^{-1}
%\nonumber\\ &=& \int\frac{d^4q}{(2\pi)^4} 
%\left[(k-q)^2-m_Q^2+i\epsilon\right]^{-1} \left[q^2-m_Q^2+i\epsilon\right]^{-1}
%\label{app:disp.2}\end{eqnarray}
 %-----------------------------------------------------------------------------------------
 In Fig.~\ref{fig:diqrk-exchange.IV} the momenta of the nucleons and quarks are shown in panel (b).
 Then, the diquark-exchange propagator in momentum space involves the integral
 \begin{eqnarray}
	 \widetilde{I}_{\mu\nu}(k;m) &=& \int\frac{d^4p}{(2\pi)^4}\ 
	 \int\frac{d^4q}{(2\pi)^4}\ (2\pi)^4\delta^4(p+q-k)
	 \left[-\left(p_\mu q_\nu +p_\nu q_\mu\right) 
	 \right.\nonumber\\ && \left.
	 +\left(p\cdot q+m^2\right) \eta_{\mu\nu}\right]\ \times 
	 \left[p^2-m^2+i\epsilon\right]^{-1} \left[q^2-m^2+i\epsilon\right]^{-1}.
 \label{app:disp.3}\end{eqnarray}
 This integral has been evaluated in \cite{BD65} exployting the Pauli-Villars
 regularization \cite{P-V.1949}, which means
 \begin{eqnarray}
	 \widetilde{I}_{\mu\nu}(k;m) &\rightarrow& \widehat{I}(k) =
	 \widetilde{I}_{\mu\nu}(k;m) + \sum_i C_i(M_i^2)\ \widetilde{I}_{\mu\nu}(k;M_i^2)
	 \equiv \sum_i c_i\ \widetilde{I}_{\mu\nu}(k;m_i^2), 
 \label{app:disp.4}\end{eqnarray}
 where the $M_i$ are large masses and the $C_i$ are chosen such that the integrals converge.
 Using the Schwinger parameterization
 \begin{eqnarray}
	 \left[p^2-m^2+i\epsilon\right]^{-1} &=&
	 -i\int_0^\infty dz\ \exp\left[iz\left(p^2-m^2+i\epsilon\right)\right]
 \label{app:disp.5}\end{eqnarray}
 the integral (\ref{app:disp.3}) takes the form
 \begin{eqnarray}
	 \widetilde{I}_{\mu\nu}(k) &=& +\int_0^\infty dz_1\ \int_0^\infty dz_2\
	 \int\frac{d^4p}{(2\pi)^4} \cdot\nonumber\\ && \times
	 \left[p_\mu(k-p)_\nu +p_\nu(k-p)_\mu -(p\cdot(k-p)+m^2)-i\epsilon)\ \eta_{\mu\nu}\right]
	 \cdot\nonumber\\ && \times \exp\left[ iz_1\left(p^2-m^2+i\epsilon\right)
	 +iz_2\left((k-p)^2-m^2+i\epsilon\right)\right]
 \label{app:disp.6}\end{eqnarray}
 Making the standard shift
 \begin{eqnarray}
	 && p_\mu \rightarrow l_\mu = p_\mu - \frac{z_2}{z_1+z_2} k_\mu =
	 (p-k)_\mu +\frac{z_1}{z_1+z_2}\ k_\mu, 
 \label{app:disp.7}\end{eqnarray}
 the denominator has $l^2$ and no linear $l_\mu$ term, which enables the integrals      
 \begin{eqnarray}
	 \int\frac{d^4p}{(2\pi)^4} \left[1, l_\mu,l_\mu l_\nu\right]\ 
	 \exp\left[i l^2(z_1+z_2)\right] = \frac{1}{16\pi^2 i}\frac{1}{(z_1+z_2)^2}
	 \left[1, 0, \frac{i\eta_{\mu\nu}}{2(z_1+z_2)}\right]
 \label{app:disp.8}\end{eqnarray}
 leading to
 \begin{eqnarray}
&& \widetilde{I}_{\mu\nu}(k) = -i \sum_i \frac{c_i}{4\pi^2}
   \int_0^\infty dz_1 \int_0^\infty \frac{dz_2}{(z_1+z_2)^2} 
 \cdot\nonumber\\ && \times
 \left(\exp\left\{i\left[k^2\frac{z_1z_2}{z_1+z_2}-(m^2-i\epsilon)(z_1+z_2)\right]\right\}\right)
 \cdot\nonumber\\ && \times
 \biggl\{ 2\left(\eta_{\mu\nu} k^2-k_\mu k_\nu\right)\frac{z_1z_2}{(z_1+z_2)^2}
 +\eta_{\mu\nu}\left[\frac{-i}{(z_1+z_2)}-\frac{k^2 z_1z_2}{(z_1+z_2)^2} +m_i^2\right]\biggr\}.
 \label{app:disp.9}\end{eqnarray}
 It appears that the $\eta_{\mu\nu}\bigl[ \cdots \bigr]$-term vanishes, see \cite{BD65}, and so
 \begin{eqnarray}
&& \widetilde{I}_{\mu\nu}(k) = -2i \sum_i \frac{c_i}{4\pi^2}
\int_0^\infty \int_0^\infty  \frac{dz_1dz_2}{(z_1+z_2)^2} \frac{z_1z_2}{(z_1+z_2)^2}\
 \cdot\nonumber\\ && \times
 \left(\exp\left\{i\left[k^2\frac{z_1z_2}{z_1+z_2}-(m_i^2-i\epsilon)(z_1+z_2)\right]\right\}\right)
 \left(\eta_{\mu\nu} k^2-k_\mu k_\nu\right)
 \label{app:disp.10}\end{eqnarray}
 Using the identity
 \begin{eqnarray}
	  1 &=& \int_0^\infty \frac{d\lambda}{\lambda}\delta\left(1-\frac{z_1+z_2}{\lambda}\right)
 \label{app:disp.10b} \end{eqnarray}
 the remaining contribution to $\widehat{I}_{\mu\nu}(k)$ becomes
 \begin{eqnarray}
	 \widehat{I}_{\mu\nu}(k) &=& -\frac{2i}{4\pi^2} \left(\eta_{\mu\nu} k^2-k_\mu k_\nu\right)
	 \int_0^\infty \int_0^\infty dz_1 dz_2 z_1 z_2 \delta(1-z_1-z_2) \int_0^\infty
	 \frac{d\lambda}{\lambda} \cdot\nonumber\\ && \hspace{3cm} \times
	 \sum_i c_i \exp\left[i\lambda \left( k^2 z_1z_2-m_i^2+i\epsilon\right)\right]
 \label{app:disp.11}\end{eqnarray}
 The $\lambda$-integral diverges logarithmically and is evaluated by applying 
 the cut-off procedure by choosing $C_1=-1, C_i=0\ (i > 1)$. This gives
 \begin{eqnarray}
	 \widehat{I}_{\mu\nu}(k) &=& 
	 \widetilde{I}_{\mu\nu}(k; m^2) - \widetilde{I}_{\mu\nu}(k; M^2) \nonumber\\ 
	 &\approx& 
	 -\frac{2i}{4\pi^2} \left(\eta_{\mu\nu} k^2-k_\mu k_\nu\right)
	 \int_0^1 dz\ z(1-z)\ \ln\left[\frac{M^2}{m^2-z(1-z)\ k^2}\right]
	 \nonumber\\ &=& 
	 -\frac{i}{12\pi^2} \left(\eta_{\mu\nu} k^2-k_\mu k_\nu\right)
	 \times \left[\ln\left(\frac{M^2}{m^2}\right) -6\int_0^1 dz\ z(1-z)\ \ln\left(1-z(1-z)\
	 \frac{k^2}{m^2}\right)\right].
 \label{app:disp.12}\end{eqnarray}
 We write $\widehat{I}_{\mu\nu}(k) \equiv 
 -i \left(\eta_{\mu\nu} k^2-k_\mu k_\nu\right)\ \Pi_2(k^2)$.
 The (unrenormalized) diquark propagator becomes
 \begin{subequations}\label{app:disp.13}
 \begin{eqnarray}
	i\widetilde{\Delta}_{\mu\nu}^{(0)}(k) &=& 
	 -\frac{i}{12\pi^2} \left(\eta_{\mu\nu} k^2-k_\mu k_\nu\right)
	 \times \left[\ln\left(\frac{M^2}{m^2}\right) -6\int_0^1 dz\ z(1-z)\ \ln\left(1-z(1-z)\
	 \frac{k^2}{m^2}\right)\right]/(\hbar c)^4 \\ 
	 & \equiv & -i\left(\eta_{\mu\nu} k^2-k_\mu k_\nu\right)\ \Pi_2(k^2)/(\hbar c)^4.
 \end{eqnarray}\end{subequations}
%--------------------------------------------------------------------------
% In the limit $k^2 \rightarrow 0$, the propagator is a constant $Z_3$, defined by
% \begin{eqnarray}
%	 Z_3 &=& 1-\frac{g_0^2}{12\pi^2} \ln\left(\frac{M^2}{m^2}\right), 
% \label{app:disp.14}\end{eqnarray}
% and the renormalized coupling is defined as
% \begin{eqnarray}
%	 g^2_R &=& Z_3 g_0^2 = 
%	  g_0^2\left[1-\frac{g_0^2}{12\pi^2} \ln\left(\frac{M^2}{m^2}\right)\right], 
% \label{app:disp.15}\end{eqnarray}
% which is independent of the momentum transfer.
%--------------------------------------------------------------------------

%--------------------------------------------------------------------------
% QQ Vacuum Polarization with Iterations
%--------------------------------------------------------------------------
%% figuur 2
%\begin{center}
% \begin{figure}[hbt]
%\centering
%%\resizebox{7.25cm}{5.75cm}
%%\resizebox{7.25cm}{6.75cm}
% \resizebox{8.25cm}{6.75cm}
% {\includegraphics[240,355][460,560]{Fig.qq-vacpol24.ps}}
%%\vspace{+0.5cm}
% \caption{Diquark-exchange NQ $\rightarrow$ QN. 
% Panel (a): axial-vector $\chi^a_\mu$-exchange. Panel (b) and (c): QQ-pair exchange and 
% iterations.}
% \label{fig:diqrk-vacpol2}
%\end{figure}
%\end{center}
%---------------------------------------------------------------------------------
% Fig.qq-vacpol24.ps 
%--------------------------------------------------------------------------
 \begin{figure}[hbtp!]
%--------------------------------------------------------------------------
 % (a) Di-quark-exchange diagram:	 
%--------------------------------------------------------------------------
 \begin{center} \begin{picture}(175,150)(75,0)
%\begin{center} \begin{picture}(175,200)(75,0)
%\begin{center} \begin{picture}(175,300)(75,0)
%\begin{center} \begin{picture}(200,175)(0,75)
%\begin{center} \begin{picture}(200,175)(0,15)
 \SetPFont{Helvetica}{9}
 \SetScale{1.0} \SetWidth{1.5}
 \SetOffset(0,0)
 \SetOffset(-70 ,0)
 \SetOffset(-140,0)
 \ArrowLine(150,15)(150,75)   
 \ArrowLine(150,75)(150,135)   
 \Vertex(150,75){3}
 \Vertex(90, 75){3}
%\DashLine(75,90)(75,150){2}
%\Photon(75,150)(75,122){1}{3}
%\Text( 75,120)[]{$\bigotimes$}
%\Photon(75,90)(75,118){1}{3}
 \ArrowLine(90,15)(90,75)   
 \ArrowLine(90,75)(90,135)   
%---------------------------------------------------------------------------
 \SetWidth{0.3}
 \SetWidth{1.0}
 \SetColor{Red}
 \Photon( 90,75)(150,75){1}{6}
%\ArrowArc(120,25 )(60,60,120)  
%\ArrowArcn(120,125)(60,-60,-120)
 \SetColor{Black}
 \SetWidth{1.5}
%\DashArrowArc(75,120)(30,-90,90){1}
%\DashArrowArcn(75,120)(30,-90,90){1}
%---------------------------------------------------------------------------
 \Text(150,0  )[]{$N,p_1$}
 \Text(150,147.5)[]{$Q,p_1'$}
 \Text(90,  0)[]{$Q,p_2$}
 \Text(90,147.5)[]{$N,p_2'$}
%\Text(120,55)[]{$p$}
%\Text(120,95)[]{$q$}
% \Text(185,75)[]{$\bigoplus$}
 \Text(175,75)[]{$=$}
 \Text(120,-5)[]{(a)}
%--------------------------------------------------------------------------
% (b) QQ Vacuum polarization diagram:
%---------------------------------------------------------------------------
  \SetOffset(140,0)
  \SetOffset(-10,0)
% \SetOffset(0,-110)
  \ArrowLine(120,15)(120,75)   
  \ArrowLine(120,75)(120,135)   
  \Vertex(120,75){3}
  \Vertex(60, 75){3}
  \Vertex(120,75){3}
  \Vertex(60, 75){3}
%\SetColor{Red}
% \Photon(180,75)(150,75){1}{3}
% \Photon(60,75)(90,75){1}{3}
  \SetColor{Black}
  \ArrowLine(60,15)(60,75)   
  \ArrowLine(60,75)(60,135)   
%---------------------------------------------------------------------------
 \SetWidth{0.5}
 \SetWidth{1.0}
%\SetColor{Red}
%\CArc(75,120)(30,-90,90)
 \SetColor{Red}
%CArc( 25,120)(60,-30,30)  
 \ArrowArc( 90,25 )(60,60,120)  
%\CArc(125,120)(60,150,210)
 \ArrowArcn( 90,125)(60,-60,-120)
 \SetColor{Black}
 \SetWidth{1.5}
%\DashArrowArc(75,120)(30,-90,90){1}
%\DashArrowArcn(75,120)(30,-90,90){1}
 \Text(145,75)[]{$\bigoplus$}
%---------------------------------------------------------------------------
 \SetWidth{1.5}
 \Text(120,0  )[]{$N,p_1$}
 \Text(120,147.5)[]{$N,p_1'$}
 \Text(60,  0)[]{$Q,p_2$}
 \Text(60,147.5)[]{$Q,p_2'$}
 \Text( 90,50)[]{$p$}
 \Text( 90,100)[]{$q$}
 \Text( 90,-5)[]{(b)}
%--------------------------------------------------------------------------
% (c) QQ Vacuum polarization iteration diagram:
%---------------------------------------------------------------------------
  \SetOffset(140,0)
  \SetOffset(100,0)
% \SetOffset(0,-110)
  \ArrowLine(180,15)(180,75)   
  \ArrowLine(180,75)(180,135)   
  \Vertex(180,75){3}
  \Vertex(120,75){3}
  \Vertex(60, 75){3}
%\SetColor{Red}
% \Photon(180,75)(150,75){1}{3}
% \Photon(60,75)(90,75){1}{3}
  \SetColor{Black}
  \ArrowLine(60,15)(60,75)   
  \ArrowLine(60,75)(60,135)   
%---------------------------------------------------------------------------
 \SetWidth{0.5}
 \SetWidth{1.0}
 \SetColor{Red}
 \ArrowArc( 90,25 )(60,60,120)  
 \ArrowArcn( 90,125)(60,-60,-120)
 \ArrowArc(150,25 )(60,60,120)  
 \ArrowArcn(150,125)(60,-60,-120)
 \SetColor{Black}
 \SetWidth{1.5}
%---------------------------------------------------------------------------
 \SetWidth{1.5}
  \Text(180,0  )[]{$N,p_1$}
  \Text(180,147.5)[]{$N,p_1'$}
  \Text(60,  0)[]{$Q,p_2$}
  \Text(60,147.5)[]{$Q,p_2'$}
%\Text( 90,50)[]{$p$}
%\Text( 90,100)[]{$q$}
 \Text(210, 75)[]{$\bigoplus\ \  \cdots$}
 \Text(120,-5)[]{(c)}
%---------------------------------------------------------------------------
 \end{picture} \end{center}
%---------------------------------------------------------------------------
%\vspace{+0.5cm}
 \caption{Diquark-exchange NQ $\rightarrow$ QN. 
 Panel (a): axial-vector $\chi^a_\mu$-exchange. Panel (b) and (c): QQ-pair exchange and 
 iterations.}
 \label{fig:diqrk-vacpol2}
\end{figure}
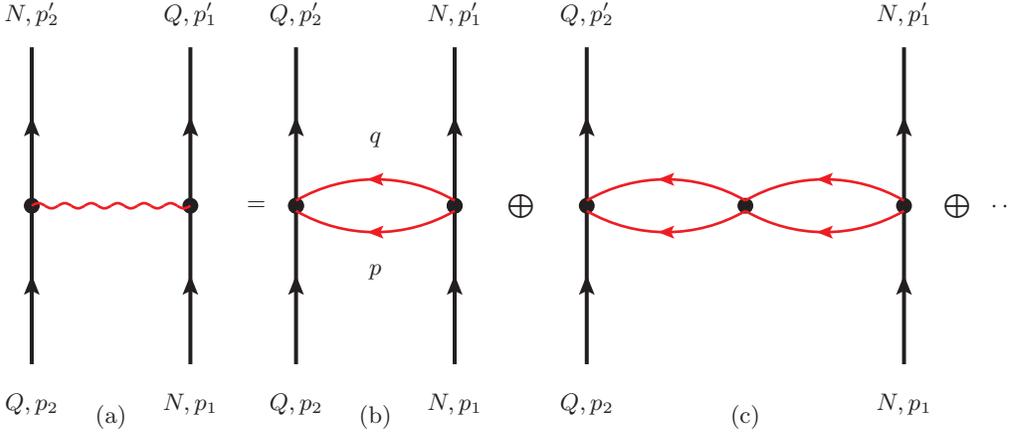
%---------------------------------------------------------------------------------

%---------------------------------------------------------------------------------
\noindent Higher-order contributions to the propagator come from diagrams as
depicted in Fig.~\ref{fig:diqrk-vacpol2}, {\it i.e.} the iteration of the 
second order diquark
contribution gives $\Pi_2(k^2) \rightarrow \Pi(k^2) = \Pi_2(k^2)/\left[1-\Pi_2(k^2)\right]$.
This is similar to that for the vacuum polarization in QED, see \cite{PS95}.
The divergency's in the diquark-propagator can be removed by a coupling constant 
renormalization, analogously to QED for the vacuum polarization.
Schematically, the matrix element for $NQ \rightarrow QN$ reads
\begin{eqnarray*}
M(k) &\sim& g_0^2\Gamma_5^\mu \Delta_{\mu\nu}(k)\ \Gamma_5^\nu \sim 
\Pi_{\mu\nu}(k^2) = \left(\eta_{\mu\nu}-k_\mu k_\nu/k^2\right)\
\Pi(k^2)\ \ {\rm with} \nonumber\\ 
\Pi(k^2) &=& \Pi_2(k^2)\left[1-\Pi_2(k^2)\right]^{-1}
\end{eqnarray*}
and considering this at $k^2=0$ the amplitude gets a factor $M(k^2=0) \sim
g_0^2/(1-\Pi_2(0)) \sim g_0^2/(1-\Pi_0(0)) \sim g_R^2$ where $g_R^2$ is the renormalized coupling.
Then, $g_0^2 = g_R^2(1-\Pi_2(0))$ 
Then, for the amplitude we get 
\begin{eqnarray*}
M(k) &\sim& g_R^2\left[\Gamma_5^\mu \left(\eta_{\mu\nu}-k_\mu k_\nu/k^2\right)\ \Gamma_5^\nu\right]\
\left\{1-\left[\Pi_2(k^2)-\Pi_2(0)\right]\right\}^{-1}
\end{eqnarray*}
where is used $(1-\Pi_2(0))/(1-\Pi_2(k^2)) \approx \left\{1-\left[\Pi_2(k^2)-\Pi_2(0)\right]\right\}^{-1}$ \cite{PS95}.                    

\noindent {\it 
%The $\chi^a_\mu$-field used thus far is not normalized to dimension [MeV]. 
%We now normalize 
%by redefining $\chi^a_\mu(x) \rightarrow \chi^a_\mu(x) (\hbar c)^2/m_\chi^2$. 
In perturbation theory the in-fields are used, 
\it i.e.} $(\Box+m_\chi^2) \chi^a_\mu(x)=0$
\footnote{Henceforth we denote the diquark mass $m_D$ by $m_\chi$.}
, 
and we take in the projection operator $k^2=m_\chi^2$.  
Then, the  propagator becomes
 \begin{subequations}\label{app:disp.16}
 \begin{eqnarray}
	i \widetilde{\Delta}_{\mu\nu}(k) &=& 
	 +\frac{2i}{4\pi^2} \left(\eta_{\mu\nu} -\frac{k_\mu k_\nu}{m_\chi^2}\right)
	 \times\left[\int_0^1 dz\ z(1-z)\ \ln\left(1-z(1-z)\
	 \frac{k^2}{m^2}\right)\right]/m_\chi^2 \\
	 &=& -\frac{i}{12\pi^2}\left(\eta_{\mu\nu}-\frac{k_\mu k_\nu}{m_\chi^2}\right)\
	 \widetilde{\Pi}(k^2)/m_\chi^2, 
 \end{eqnarray}\end{subequations}
 where $\widetilde{\Pi}(k^2)= \Pi_2(k^2)-\Pi_2(0)$.
 The z-integral is elementary and can be found in the literature, 
 see {\it e.g.} \cite{Kallen52,J-R.1976},
%G.\ K\"{a}ll\'{e}n, {\it Handbuch der Physik}, Vol. V- Part I, equation (29.35), and 
%\cite{J-R.1976} J.M.\ Jauch and F.\ Rohrlich, 
%{\it the Theory of Photons and Electrons}, equation (A5-27), 
 and with $A=m^2/k^2$ reads
 \begin{eqnarray}
	 && \int_0^1 dz\ z(1-z)\ \ln\left(1-z(1-z)\
	 \frac{k^2}{m^2}\right) =
  \frac{1}{6}\biggl\{ (1+2A) \sqrt{1-4A}\ 
	 \ln\left[\frac{1+\sqrt{1-4A}}{\left|1-\sqrt{1-4A}\right|}\right]
 -\left(4A+\frac{5}{3}\right)\biggr\}.     
 \label{app:disp.17}\end{eqnarray}
 It can be verified that for $k^2 \rightarrow 0$ the -4A-term in canceled, and there
 is no pole.
 %-------------------------------------------------------------------
% %-------------------------------------------------------------------
% \noindent  The x-integral in $\widetilde{I}_2$ has the form 
% \begin{eqnarray*}
% && \int_0^1 x^2dx\ \ln\left[(x-x_+)(x-x_-)\right] =
% \int_0^1 x^2dx\ \left[\ln(x-x_+)+\ln(x-x_-)\right].  
% \end{eqnarray*}
% %-------------------------------------------------------------------
% %-------------------------------------------------------------------
% Via partial integration 
% and using $x^3 = (x-a)^3+3ax(x-a)+a^3$, we obtain
% \begin{eqnarray*}
%	 \int_0^1 x^2dx\ \ln(x-a) &=& \frac{1}{3}\left[(1-a^3) \ln(1-a) +a^3 \ln(-a)
%	 -\left(a^2+\frac{1}{2}a + \frac{1}{3}\right) \right].
% \end{eqnarray*}
% and analogously
% \begin{eqnarray*}
%	 \int_0^1 xdx\ \ln(x-a) &=& \frac{1}{2}\left[(1-a^2) \ln(1-a) +a^2 \ln(-a)
%	 -\frac{1}{2}\left(2a + 1\right) \right].
% \end{eqnarray*}
%
% This gives
% \begin{eqnarray*}
% && \int_0^1 xdx\ \ln\left[(x-x_+)(x-x_-)\right]= \frac{1}{2}\left[
% \sqrt{1-4A}\ \ln\frac{1+\sqrt{1-4A}}{1-\sqrt{1-4A}}
%	 +\ln A -2 \right], \\
% && \int_0^1 x^2dx\ \ln\left[(x-x_+)(x-x_-)\right] =
% \frac{1}{3}\bigg[(1-A)\sqrt{1-4A}\ \ln\frac{1+\sqrt{1-4A}}{1-\sqrt{1-4A}} 
% +\ln A+(2A-\frac{13}{6})\bigg], \nonumber\\ 
% && \int_0^1 dx\ x(1-x)\ \ln\left[(x-x_+)(x-x_-)\right]= \frac{1}{6}\left[
%	 (1+2A) \sqrt{1-4A}\ \ln\frac{1+\sqrt{1-4A}}{1-\sqrt{1-4A}}
%	 +\ln A -(4A+\frac{5}{3}) \right]
% \end{eqnarray*}
% %-------------------------------------------------------------------
 The discontinuity in the complex k-plane is 
 \begin{eqnarray}
% Disc\ \widetilde{I}_2(k) &=& \frac{2\pi i}{3}\ 
% \left(1-\frac{m^2}{k^2}\right) \sqrt{1-\frac{4m^2}{k^2}}\ 
% \theta\left(1-\frac{4m^2}{k^2}\right), \\
% Disc\ \widetilde{I}_1(k) &=& \frac{2\pi i}{4}\ 
% \sqrt{1-\frac{4m^2}{k^2}} \sqrt{1-\frac{4m^2}{k^2}}\ 
% \theta\left(1-\frac{4m^2}{k^2}\right), \\
 Disc\ \widetilde{I}(k) &=& \frac{2\pi i}{6}\ 
 (1+\frac{2m^2}{k^2}) \sqrt{1-\frac{4m^2}{k^2}}\ 
	 \theta\left(1-\frac{4m^2}{k^2}\right)
 \label{app:disp.19}\end{eqnarray}
 Then, in momentum space, using a cut-off $s_{max}$, 
 \begin{eqnarray}
	i (\widetilde{\Delta}_F)_{\mu\nu}(k) &=& 
	 +\frac{i}{12\pi^2} \left(\eta_{\mu\nu} -\frac{k_\mu k_\nu}{m_\chi^2}\right)
	 \times
	 (\hbar c)^{-2} \int_{4m^2}^{s_{max}} ds\
	\left(1+\frac{2m^2}{s}\right)\sqrt{1-\frac{4m^2}{s}} 
	\bigl[k^2-s+i\epsilon\bigr]^{-1}.
 \label{app:disp.20}\end{eqnarray}
 %-----------------------------------------------------------------------------------------
 Separation the finite and divergent part is achieved by using \cite{Kallen52}
 \begin{eqnarray*}
	 && \frac{1}{k^2-s+i\epsilon} = \frac{k^2}{s(k^2-s+i\epsilon)} -\frac{1}{s-i\epsilon}
 \end{eqnarray*}
 which gives for the finite part, {\it i.e.} the "renormalized propagator" 
 \footnote{This dispersive method of "renormalization" is from \cite{Kallen52}, and is an
 alternative to the Pauli-Villars method.},
 \begin{eqnarray}
	i (\widetilde{\Delta}_F)_{\mu\nu}(k) &=& 
	 +\frac{i}{12\pi^2} \left(\eta_{\mu\nu} -\frac{k_\mu k_\nu}{m_\chi^2}\right)
	 \times
	 \frac{k^2}{(\hbar c)^2} \int_{4m^2}^{\infty} ds\
	\left(1+\frac{2m^2}{s}\right)\sqrt{1-\frac{4m^2}{s}} 
	 \bigl[s\left(k^2-s+i\epsilon\right)\bigr]^{-1}.
 \label{app:disp.201}\end{eqnarray}
 In the process $N \rightarrow Q+D$ we approximate 
 $k^2 \approx k_0^2 \approx (m_N-m_Q)^2 \equiv (\Delta m_{NQ})^2$   
 valid for low-momentum transfer.\\
 %-----------------------------------------------------------------------------------------
 The spectral representation of the Diquark Feynman propagator is 
 \begin{eqnarray}
	 i(\Delta_F)_{\mu\nu}^{ab}(x'-x) &=& 
	 (0|T\left[\chi^a_\mu(x') \chi^{b \dagger}_\nu(x)\right]|0)
	 = i \left(\frac{\Delta m_{NQ}}{\hbar c}\right)^2
	 \int_{4m^2}^{s_{max}} ds\ (\Delta_F)^{ab}_{\mu\nu}(x'-x;s)\ \rho(s)
	 \nonumber\\ &=& -16i\ \left(\frac{\Delta m_{NQ}}{\hbar c}\right)^2
	 \delta_{ab}\ D(x'-x;m_\chi,\Lambda)\
	 \left[\eta_{\mu\nu}-\frac{k_\mu k_\nu}{m_\chi^2}\right].
 \label{app:disp.31}\end{eqnarray}
 %-----------------------------------------------------------------------------------------
 The Fourier transforms, with $m_\chi \sim 2 m_Q$, are
 \begin{subequations}\label{app:disp.32}
 \begin{eqnarray}
 \widetilde{D}(k^2,m_\chi^2) &=& 
	 - \left(\frac{\Delta m_{NQ}}{\hbar c}\right)^2
	 \int_{4m_Q^2}^{s_{max}} ds\ \rho(s)\ \widetilde{\Delta}_F(k^2,\Lambda^2;s), \\
 \widetilde{\Delta}_F(k^2;s) &=& \left[k^2-s+i\epsilon \right]^{-1},\ \
 \rho(s) = \frac{1}{12\pi^2} \left(1+\frac{2m^2}{s}\right) \sqrt{1-\frac{4m^2}{s}}/s.
 \end{eqnarray}\end{subequations}
	 With $\sqrt{s} = E({\bf q})= \sqrt{{\bf q}^2+m_\chi^2}$ 
	 we have for space-like $k^2=-{\bf k}^2$ 
 \begin{subequations}\label{app:disp.33}
 \begin{eqnarray}
%\widetilde{\Delta}_F({\bf k}^2,\Lambda^2;s) &=&+\left[{\bf k}^2+{\bf q}^2+m_\chi^2\right]^{-1}, \ \
	 \widetilde{D}({\bf k}^2,m_\chi^2) &=&
	 -4\pi  \left(\frac{\Delta m_{NQ}}{\hbar c}\right)^2
	 \int_{0}^\infty q^2dq\ \rho(s)\ {\Delta}_F({\bf k}^2,\Lambda^2;s), \\    
	 \widetilde{\Delta}_F({\bf k}^2,\Lambda^2;s) &=& -\exp\left[-{\bf k}^2/\Lambda^2\right]
	 \left[{\bf k}^2+{\bf q}^2+m_\chi^2\right]^{-1}, 
%\label{app:disp.33} \end{eqnarray}
 \end{eqnarray}\end{subequations}
 where we added a gaussian form-factor as usual in the Nijmegen potentials.
%-----------------------------------------------------------------------------------------
 In configuration space
 \begin{eqnarray}
	 D({\bf x}^2,m_Q^2) &=& \left(\frac{\Delta m_{NQ}}{\hbar c}\right)^2
	 \int_{4m_Q^2}^{s_{max}} ds\ \rho(s)
 \left[\frac{m(s)}{4\pi}\ \phi_C^0\left(m(s),\Lambda; |{\bf x}|\right)\right]
 \label{app:disp.34}\end{eqnarray}
 where $m(s)= \sqrt{s-m_\chi^2}$.\\

 %-----------------------------------------------------------------------------------------
 {\bf This result implies a repulsive QN-potential!}

%--------------------------------------------------------------------------
 \begin{center}
	 \fbox{ \begin{minipage}[b]{16cm}
 \vspace{2mm}
%--------------------------------------------------------------------------
 \begin{center}
 \underline{\blue Gaussian Approximation}:    
 \end{center}
 The volume integral $D_V$ of the D(x)-function is
 \begin{eqnarray}
 D_V &=& \left(1+2\frac{m_Q}{\Delta m_{NQ}}\right)\
 \int_{m_Q^2}^\infty ds\ \rho(s)/m^2(s). 
 \label{app:disp.21}\end{eqnarray}
 where we included a factor which accounts for the mass-difference at the NQ-vertex giving
		 $k_0 \rightarrow m_n-m_Q$. 
 For a gaussian approximation 
 \begin{equation}
 D_G(|{\bf x}|) = g \exp\left[-|{\bf x}|^2/\Lambda^2\right],\ \
 \label{app:disp.22}\end{equation}
 having the same volume integral one has
 \begin{equation}
 2\pi\sqrt{\pi}\ g\Lambda^3 = \left(1+2\frac{m_Q}{\Delta m_{NQ}}\right)\
\int_{m_Q^2}^\infty ds\ \rho(s)/m^2(s). 
 \label{app:disp.23}\end{equation}

 \end{minipage} }\\
 \end{center}
%--------------------------------------------------------------------------
%------------------------------------------------------------------
\begin{flushleft}
\rule{16cm}{0.5mm}
\end{flushleft}
%------------------------------------------------------------------
\section{ Di-quark exchange Nucleon-Quark Interaction}
\label{sec:diquarkpot}  
%--------------------------------------------------------------------------
As the proton and neutron presentation \cite{Ioffe81} shows the diquark field has isospin one. So, 
$\chi_\mu^a(x)$ is an isovector vector-field. Therefore, we introduce the fields
\begin{equation}
	{\bf D}^a_{\mu}(x) \equiv 
	\bm{\chi}_\mu^a(x) = \varepsilon^{abc}\tilde{Q}^b(x) C\gamma_\mu \bm{\tau}Q^c(x)/(\hbar c)^2,
\label{eq:Mix:9}\end{equation}
where  $Q=(u,d)$  is the isospin-spinor SU$_I$(2) doublet and a SU$_C$(3) triplet. Similarly   
for the di-quark ${\bf D}^a_\mu$.         
The di-quark NQ-vertex is given by the interaction Lagrangian
\begin{equation}
	{\cal L}_{int}^{(1)} = -\lambda_3 \bigl\{(\bar{\psi}(x)\gamma_5\gamma^\mu 
	\bm{\tau}q^a)\cdot{\bf D}_\mu^a(x) + {\it h.c.} \bigr\},
\label{eq:Mix:10}\end{equation}
%--------------------------------------------------------------------------
 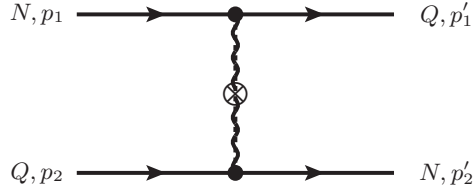
\begin{figure}[hbtp!]
%--------------------------------------------------------------------------
% axodraw figure:     
%\begin{center} \begin{picture}(350,175)(0,75)
 \begin{center} \begin{picture}(200,175)(0,75)
 \SetPFont{Helvetica}{9}
 \SetScale{1.0} \SetWidth{1.5}
 \SetOffset(0,0)
 \ArrowLine(15,150)(75,150)   
 \ArrowLine(75,150)(135,150)   
 \Vertex(75,150){3}
 \Vertex(75, 90){3}
 \DashLine(75,90)(75,150){2}
 \Photon(75,150)(75,122){1}{3}
 \Text( 75,120)[]{$\bigotimes$}
 \Photon(75,90)(75,118){1}{3}
 \ArrowLine(15,90)(75,90)   
 \ArrowLine(75,90)(135,90)   
 \Text( 0, 150)[]{$N,p_1$}
 \Text(155,150)[]{$Q,p_1'$}
 \Text( 0, 90)[]{$Q,p_2$}
 \Text(155,90)[]{$N,p_2'$}
%\Text(75, 75)[]{(a)}
%---------------------------------------------------------------------------
 \end{picture} \end{center}
% \vspace{-5mm}
%\caption{Diquark-exchange (a) $NQ \rightarrow QN$ and                       
%(b) $NQ \rightarrow NQ$ transitions.}             
 \caption{Diquark-exchange  for $NQ \rightarrow QN$ reaction.}                         
  \label{fig:trans.1a}
\end{figure}
%---------------------------------------------------------------------------------
%------------------------------------------------------------------------------
%% figuur 2
%\begin{center}
% \begin{figure}[hbtp!]
%\centering
%%{\includegraphics*[width=16cm,height=21cm]{./Fig.triquark-nucleon.ps}}
%%\caption{Di-quark-quark--Nucleon Vertex}                                            
% \resizebox{7.25cm}{5.75cm}
%%\resizebox{6.25cm}{2.75cm}
%%{\includegraphics*[width=16cm,height=8cm]{./Fig.diquark-quark-nucleon.ps}}
% {\includegraphics[180,475][400,655]{Fig.diquark-NQQN.ps}}
% \caption{Diquark-exchange  for $NQ \rightarrow QN$ reaction.}                         
%  \label{fig:trans.1a}
%\end{figure}
%\end{center}
%------------------------------------------------------------------------------
 The field theoretical study of the diquark-propagator in this paper shows that 
 the Feynman-rule for the diquark $\chi^a$-propagator for the $\eta_{\mu\nu}$-term is
 found in (\ref{app:disp.10})
 \begin{equation}
	 i(\widetilde{\Delta_F})_{\mu\nu}^{ab}(\bm{\Delta}) = 
	 -i \delta_{ab} \widetilde{D}(\bm{\Delta}^2)\ 
	 \left(\eta_{\mu\nu}-\frac{\bm{\Delta}_\mu \bm{\Delta}_\nu}{m_\chi^2}\right),\ \ 
	 \widetilde{D}(\bm{\Delta}^2) >0,
\label{eq:Mix:11}\end{equation}
where $\Delta = p_1'-p_1=p_2'-p_2$. 
In the gaussian approximation
 \begin{equation}
	 \widetilde{D}(\bm{\Delta}^2) \approx exp\left[-\bm{\Delta}^2/\Lambda^2\right]/{\cal M}^2.
\label{eq:Mix:111}\end{equation}

%---------------------------------------------------------------------------------
\noindent The second-order amplitude for Fig.~\ref{fig:trans.1a} using the interaction (\ref{eq:Mix:10})   
can be described (effectively) by 
 \begin{eqnarray}
 M^{(2)}(p_1',s_1',p_2',s_2'; p_1,s_1,p_2,s_2) &=& -\frac{\lambda_3^2}{2!}\ 
\left[\bar{u}_Q(p_1',s_1') \gamma_5\gamma^\mu \bm{\tau}  u_N(p_1,s_1)\right]\cdot
\left[\bar{u}_N(p_2',s_2') \gamma_5\gamma_\mu \bm{\tau}  u_Q(p_2,s_2)\right]\ 
	 \widetilde{D}(\bm{\Delta}^2),
%\cdot\nonumber\\ && \times \left(\Delta^2-m_\chi^2+i\delta \right)^{-1},
\label{eq:Diqrk.2aa}\end{eqnarray}
where $\Delta = p_1'-p_1=p_2'-p_2$. 
%-----------------------------------------------------------------
 In the low momentum transfer region the approximation 
	 $\widetilde{D} \approx \exp\left[-\bm{\Delta}^2/\Lambda^2\right]/{\cal M}^2$ leads to
a gaussian contact interaction.                                  
%-----------------------------------------------------------------
\begin{eqnarray}
	V(p_1',s_1',p_2',s_2'; p_1,s_1,p_2,s_2) &=& -(\lambda_3^2/2) 
	\left[\bar{u}_Q(p_1',s_1') \gamma_5\gamma^\mu \bm{\tau}  u_N(p_1,s_1)\right]\cdot
	\left[\bar{u}_N(p_2',s_2') \gamma_5\gamma_\mu \bm{\tau}  u_Q(p_2,s_2)\right]\ 
	\widetilde{D}(\bm{\Delta}^2).
\label{eq:Diqrk.2a}\end{eqnarray}

\noindent Using Pauli-spinor matrix elements
\begin{eqnarray}
	\bar{u}({\bf p}'\gamma_5\gamma_0 u({\bf p}) &=& -\sqrt{\frac{{\cal E'}{\cal E}}{4M'M}}
	\left[\frac{\bm{\sigma}\cdot{\bf p}'}{{\cal E}'} +
	      \frac{\bm{\sigma}\cdot{\bf p}}{{\cal E}}\right], \\
	\bar{u}({\bf p}')\gamma_5\bm{\gamma} u({\bf p}) &=& -\sqrt{\frac{{\cal E'}{\cal E}}{4M'M}}
	\left[ \bm{\sigma} + \frac{(\bm{\sigma}\cdot{\bf p}')\ \bm{\sigma}\ 
	(\bm{\sigma}\cdot{\bf p}) }{{\cal E}'{\cal E}}\right] \approx -\bm{\sigma}, 
\label{eq:Diqrk.2b}\end{eqnarray}
where M',M are the quark or the nucleon mass, and ${\cal E} = E_p+M$.
Note that the leading term from the vertex factors [....][....] in (\ref{eq:Diqrk.2a})  
has $-(\bm{\sigma}_1\cdot\bm{\sigma}_2)$.
%{\red second (-)-sign w.r.t. L_{int}, --> V_{QN} same sign as L_{int}!!}
In momentum space we write $\widetilde{V}_{QN} \equiv \widetilde{V}_{QN}^{(a)}+\widetilde{V}_{QN}^{(b)}$,
where (a) is similar to the axial-exchange in NN-potentials, and (b) represents terms emphasizing the nucleon and 
quark mass, and obtain 
%{\red version 5 may 2023 has in k^2-term 1/3 instead of 2/3!}
\begin{subequations}\label{eq:Diqrk.2c}
\begin{eqnarray}
	\widetilde{V}_{QN}^{(a)} &=& +2\lambda_3^2 \biggl[ \left(
	1-\frac{2{\bf k}^2}{3M_QM_N} + \frac{3({\bf q}^2+{\bf k}^2/4)}{2M_QM_N}\right)\
	\bm{\sigma}_1\cdot\bm{\sigma}_2 
	+\frac{1}{4M_QM_N}\left(
	(\bm{\sigma}_1\cdot{\bf k}) (\bm{\sigma}_2\cdot{\bf k}) -\frac{1}{3}{\bf k}^2
	\bm{\sigma}_1\cdot\bm{\sigma}_2 \right)                        
	\nonumber\\ && \hspace{1cm} 
        +\frac{i}{4M_QM_N} (\bm{\sigma}_1+\bm{\sigma}_2)\cdot {\bf q}\times{\bf k}\biggr]
	\cdot \widetilde{g}({\bf k}^2), \\
%\label{eq:Diqrk.2c}\end{eqnarray}
%\begin{eqnarray}
	\widetilde{V}_{QN}^{(b)} &=& -2\lambda_3^2 \biggl[ 
	\frac{(M_N-M_Q)^2}{4M_N^2M_Q^2}\bigl\{ \left({\bf q}^2+{\bf k}^2/4\right)-{\bf k}^2/2\bigr\}\
	\bm{\sigma}_1\cdot\bm{\sigma}_2 \nonumber\\ &&
	-\frac{i}{4}\left(\frac{M_N^2-M_Q^2}{8M_N^2M_Q^2}\right)\ 
	\left(1+\bm{\sigma}_1\cdot\bm{\sigma}_2\right)
	\left(\bm{\sigma}_1-\bm{\sigma}_2\right)\cdot{\bf n} \biggr] 
	\cdot \widetilde{g}({\bf k}^2), 
\end{eqnarray}\end{subequations}
where $\widetilde{g}({\bf k}^2) = \exp\left(-{\bf k}^2/\Lambda^2\right)/{\cal M}^2$.
Here, we added the gaussian cut-off and a scale parameter ${\cal M}$.\\
	{\it Note: $V_{QN}^{(a)}$ is similar to axial-vector exchange in NN and YN.
	$V_{QN}^{(b)}$ is the "extra term" proportional to the $M_N-M_Q$ mass difference,
	which is not small in the QN-potential.}

%--------------------------------------------------------------------------
 \vspace{3mm}
\begin{center}
\hspace{-1cm}\fbox{ \begin{minipage}[b][15.5cm][l]{17.5cm}
\vspace{2mm}
%--------------------------------------------------------------------------
\noindent In configuration space, taking into account the exchange character of the potential 
we have a factor $P_fP_\sigma$. Since the physical states satisfy $P_fP_\sigma P_x=-1$,
this leads to a factor $-P_x$ and a sign-change in the antisymmetric spin-orbit. 
%--------------------------------------------------------------------------
 Then, we obtain for the central, spin-spin, tensor, and
	spin-orbit-potentials, see {\it e.g.} Ref.~\cite{RNY2010}, 
%------------------------------------------------------------------
%\noindent {\red CHECK SIGN!!!}
\begin{subequations}\label{eq:Diqrk.4} 
\begin{eqnarray}
	V_{QN}^{(a)}(r) &=& -2\lambda_3^2\ \frac{\Lambda}{8\pi}\ \biggl[
	\left(\phi_C^0(r)-\frac{\Lambda^2}{6M_NM_Q} \phi_C^1(r)\right) 
(\bm{\sigma}_1\cdot\bm{\sigma}_2) 
	-\frac{3}{4M_QM_N}\left(\bm{\nabla}^2\phi_C^0(r)+\phi_C^0(r)\bm{\nabla}^2\right)
	(\bm{\sigma}_1\cdot\bm{\sigma}_2) \nonumber\\
&& -\frac{\Lambda^2}{16M_NM_Q}\ \phi_T^0(r)\ S_{12}
+\frac{\Lambda^2}{8M_NM_Q} \phi_{SO}^0(r)\ {\bf L}\cdot{\bf S} 
\biggr]\ (\bm{\tau}_1\cdot\bm{\tau}_2)\ P_x, \\
%------------------------------------------------------------------
%\begin{subequations}\label{eq:Diqrk.4a} 
%\begin{eqnarray}
	V^{(b)}_{QN}(r) &=& -2\lambda_3^2\ \frac{\Lambda}{8\pi}\biggl[
		\frac{(M_N-M_Q)^2}{4M_NM_Q} \biggl\{ +\frac{\Lambda^2}{8M_NM_Q} \phi_C^1(r)  
	 +\frac{1}{2M_NM_Q}\left(\bm{\nabla}^2\phi_C^0(r)+\phi_C^0(r)\bm{\nabla}^2\right)  \biggr\}\ 
	 \cdot\nonumber\\ && \times
	(\bm{\sigma}_1\cdot\bm{\sigma}_2)\ 
	-\frac{\Lambda^2}{4M_NM_Q} \frac{(M_N^2-M_Q^2)}{4M_N M_Q}
	\phi_{SO}^0(r)\cdot
	\frac{1}{2} (\bm{\sigma}_1-\bm{\sigma}_2)\cdot{\bf L}\biggr]\ 
	(\bm{\tau}_1\cdot\bm{\tau}_2)\ P_x,
\end{eqnarray} \end{subequations}
%%------------------------------------------------------------------
%%------------------------------------------------------------------
where 
\begin{subequations}\label{eq:Diqrk.5} \begin{eqnarray}
\phi_C^0(r) &=& \frac{1}{\sqrt{\pi}}\frac{\Lambda^2}{{\cal M}^2}\ 
	\exp\left[-\frac{1}{4}\Lambda^2r^2\right], \\
\phi_C^1(r) &=& \frac{2}{\sqrt{\pi}}\frac{\Lambda^2}{{\cal M}^2}\ 
	\left(3-\Lambda^2r^2/2\right)\exp\left[-\frac{1}{4}\Lambda^2r^2\right], \\
\phi_T^0(r) &=& \frac{1}{6\sqrt{\pi}}\frac{\Lambda^2}{{\cal M}^2}\ 
	(\Lambda r)^2 \exp\left[-\frac{1}{4}\Lambda^2r^2\right], \\
\phi_{SO}^0(r) &=& \frac{2}{\sqrt{\pi}}\frac{\Lambda^2}{{\cal M}^2}\ 
	\exp\left[-\frac{1}{4}\Lambda^2r^2\right]. 
\end{eqnarray}\end{subequations}
We introduced a gaussian cut-off with the parameter $\Lambda$. This parameter
is a free parameter and can be used to tune the di-quark exchange potential which is
also the case with $\lambda_3$.
The non-local potential is
\begin{eqnarray}
	V^{(n.l.)}(r) &=& -\left[\bm{\nabla}^2\frac{\phi(r)}{2M_{red}} 
+\frac{\phi(r)}{2M_{red}}\bm{\nabla}^2\right] P_x,\ \rm{with}\ 
	\phi(r)= -\frac{\lambda_3^2}{4\pi}\frac{3\Lambda}{4M_QM_N}\phi_C^0(r)\ 
	(\bm{\sigma}_1\cdot\bm{\sigma}_2)\
	(\bm{\tau}_1\cdot\bm{\tau}_2).
\label{eq:Diqrk.6}\end{eqnarray}

\end{minipage} }\\
\end{center}
 \vspace{2mm}
%------------------------------------------------------------------------------

%------------------------------------------------------------------
For the statistical average S-wave potential we obtain from Eq.~(\ref{eq:Diqrk.4})
\begin{eqnarray}
\bar{V}(CQM) &=& \frac{1}{4}V(^1S_0)+\frac{3}{4}V(^3S_1) 
= +\frac{3\lambda_3^2}{4\pi}\ \Lambda 
\left(\phi_C^0(r)-\frac{\Lambda^2}{6M_NM_Q}\left\{1-\frac{3(M_N-M_Q)^2}{16 M_NM_Q}\right\}
		\phi_C^1(r)\right) 
\label{eq:Diqrk.7} \end{eqnarray}
which result comes from                   
$(\bm{\sigma}_1\cdot\bm{\sigma}_2)
 (\bm{\tau}_1\cdot\bm{\tau}_2)= -3$ for both $^1S_0$ and $^3S_1$.\\

\noindent The confinement-deconfinement transition can be parametrized as  
$\lambda_3 \rightarrow \gamma_D\lambda_3$ with {\it e.g.} 
\begin{eqnarray}
\gamma_D(\rho_N,\rho_D) &=& \left[\exp\bigl\{+\gamma_3
\left(\rho_N/\rho_D-1\right)\bigr\}-1\right]\ \theta(\rho_N-\rho_D), 
\label{eq:Diqrk.8}\end{eqnarray}
where $\rho_D$ is the deconfinement threshold. 
In \cite{YYR22,YYR23,YYR24} a similar form is used for the density dependence of 
the constituent quark mass.

\noindent Notes: {\it 1. The  S-wave quark-nucleon repulsion 
(\ref{eq:Diqrk.7}) is repulsive and becomes strong for high densities.     
2. The $^1P_1$-wave has $ \bm{\sigma}_1\cdot\bm{\sigma}_2  
\bm{\tau}_1\cdot\bm{\tau}_2 P_x= -9$ giving strong repulsion. For $^3P_J (J=0,1,2)$ 
the spin-isospin and the exchange operator give a factor -1, giving again a
(weaker) repulsion. 
3. The di-quark exchange potential gives a repulsive wall for the nucleons
between the nucleon- and quark-phase.
}\\

%\noindent Below, we give an illustration of the QN potentials for three values of the
%gaussian cut-off parameter: $\Lambda= 300, 542$, and 700 MeV, which lead to
%three different types of curves.
%\noindent In Fig.~\ref{fig:diquarkpota} the S- and P-wave di=quark exchange quark-nucleon
%potentials are shown. The parameters are: $\lambda_3 =5.4, m_\chi= 2m_Q$.
%(Note that $V(^3S_1)=V(^1S_0)$, so there is in fact only one S-wave potential.)

%--------------------------------------------------------------------------
\section{Discussion and Conclusion} \label{conclusion}
%--------------------------------------------------------------------------
 For antiquark-quark$(\bar{Q}Q)$ exchange, which is the case for {\it e.g.} $\rho$ and $A_{1}$,
 there is a (-)-sign coming from the closed fermion loop. For diquark(QQ) exchange this (-)-sign
 is absent. 
 This follows from the analysis of the Wick-expansion of the T-product
 $(0|T\left[\chi^{(a)}_\mu(x) \chi^{b \dagger}_\nu(x')\right]|0)$, and explains the 
 sign difference between the $\rho,A_1$ propagators and the D-propagator.
%--------------------------------------------------------------------------
%The difference between diquark-exchange and quark-antiquark-exchange, in the case of for example
%$\rho, A_1$-exchange, is the absence of the (-)-sign due to the closed fermion-loop for
%diquark-exchange. This is clear from the analysis of the Wick-expansion of the T-product
%$(0|T\left[\chi^{(a)}_\mu(x) \chi^{b \dagger}_\nu(x')\right]|0)$.
%--------------------------------------------------------------------------
Therefore, in the Feynman-rule for diquark-exchange there is the factor $+iP_{\mu\nu}$
instead of $-iP_{\mu\nu}$. 
This, and the exchange-character of the $QN \rightarrow NQ$ interaction, 
is the source of the repulsion of the diquark-exchange
with the $\gamma_5\gamma_\mu$-coupling. The diquark is in the $\{\bar{3}\}$-irrep of SU$_c$(3),
which gives a factor $C_F=2$.\\
\noindent We calculate the diquark-exchange $NQ \rightarrow QN$ potential using the effective propagator
\begin{eqnarray*}
i(\widetilde{\Delta})^{ab}_{\mu\nu}(k) &=& 
	+2i\delta_{ab}\frac{\left(\eta_{\mu\nu}-k_\mu k_\nu/m_\chi^2\right)} 
	{k^2-m_\chi^2+i\epsilon},
\end{eqnarray*}
where $m_\chi$ is an effective mass $\approx 2m_Q$.
The vector and axial-vector mesons $\rho$ and $A_1$  
show up as resonances $\pi\pi$ and $\pi\rho$ channels. In the quark-model they are antiquark-quark
systems, and a calculation similar to that for the qq-system in this paper applies to them as well.
In this view also the $\rho$ and $A_1$ propagators are effective ones just as the diquark D propagator.\\
%--------------------------------------------------------------------------
%We calculate the diquark-exchange $NQ \rightarrow QN$ potential using the effective propagator
%\begin{eqnarray*}
%i(\widetilde{\Delta})^{ab}_{\mu\nu}(k) &=& 
%	+2i\delta_{ab}\frac{\left(\eta_{\mu\nu}-k_\mu k_\nu/m_\chi^2\right)} 
%	{k^2-m_\chi^2+i\epsilon},
%\end{eqnarray*}
%where $m_\chi$ is an effective mass $\approx 2m_Q$.
%We note the (-)-sign difference with a $\rho$ and $A_1$ propagator which is explained in these notes.
%--------------------------------------------------------------------------
{\it
\noindent The derivation in these notes establishes without any ambiguity the repulsive 
quark-nucleon (QN) for the axial-vector diquark-exchange potential, 
which is used in \cite{YYR24} for nuclear-quark mixed matter 
and apllied succesfully to heavy neutron stars $M_{NS} \approx 2.1 M_{\bigodot}$.
}\\
\noindent A second proton current $\eta^{(2)}(x)$ is treated in Appendix~\ref{app:scal.a} and 
 contains a scalar $\chi^a_S(x)$ and a pseudoscalar $\chi^a_5(x)$ diquark field, which 
introduces extra $NDQ$-vertices. They lead to a scalar and pseudoscalar NQ-potential. The scalar 
field allows the appearance of a condensate $\langle 0| \chi_S | 0\rangle \ne 0$ \cite{ARW98}, a
possibility studied extensively in the literature. Whereas for the axial-vector diquark-exchange 
we found a repulsion in all partial waves, for the scalar-pseudoscalar diquarks
the QN-potentials are a mixture of attraction and repulsion.

%--------------------------------------------------------------------------
\appendix

%--------------------------------------------------------------------------
 \section{Diquark Feynman-propagator and Wick-expansion}
 \label{app:FD-WE}
%--------------------------------------------------------------------------
% \begin{center}
%	 \fbox{ \begin{minipage}[b]{16cm}
% \vspace{2mm}
%--------------------------------------------------------------------------
% \begin{center}
%	 {\blue \underline{Diquark Feynman-propagator}:}
% \end{center}
%------------------------------------------------------------------------------
 The diquark field Feynman propagator is 
 \begin{eqnarray}
 i(\Delta_F)_{\mu\nu}^{ab}(x'-x) &=& (0|T\left[\chi^a_\mu(x') \chi^{b \dagger}_\nu(x)\right]|0)
 \label{app:propa31}\end{eqnarray}
where the diquark fields are
\begin{eqnarray}
\chi_\mu^a(x) &=& \widetilde{q}^b(x) C\gamma_\mu q^c(x)\ \varepsilon^{abc}\ ,\ 
\chi_\mu^{a \dagger}(x) = -\bar{q}^b(x) \gamma_\mu C\ \widetilde{\bar{q}}^c(x)\ \varepsilon^{abc}
 \label{app:propa32}\end{eqnarray}
 The Wick-expansion of the T-product into normal-ordered N-products for the operators
 A,B,C, and D occurring in (\ref{app:propa31}) reads, in the notation of \cite{Mandl61},
 \begin{eqnarray*}
	 T(ABCD) &=& N(ABCD) + N(\widehat{AB}\ CD) + N(AB\ \widehat{CD}) + N(\widehat{AB}\ \widehat{CD})
	 -N(\widehat{AC}\ \widehat{BD}) + N(\widehat{AD}\ \widehat{BC}) \nonumber\\ & \Rightarrow&
	 -N(\widehat{AC}\ \widehat{BD}) + N(\widehat{AD}\ \widehat{BC}),
 \end{eqnarray*}
 where the terms in the last line survives after taking the vacuum expectation value.
		 In the notation of \cite{BD65} this reads
 \begin{eqnarray*}
 T(ABCD) &=& :ABCD: + \langle 0|T(AB)|0\rangle :CD: + :AB:\langle 0|T(CD)|0\rangle + 
	 \langle 0|T\rangle(AB)|0\rangle \langle |T(CD)|0\rangle  
 \nonumber\\ && 
 -\langle 0|T(AC)|0\rangle  \langle 0|T(BD)|0\rangle  + \langle 0|T(AD)|0\rangle  \langle 0|T(BC)|0\rangle 
 \nonumber\\ & \Rightarrow&
 -\langle 0|T(AC)|0\rangle  \langle 0|T(BD)|0\rangle  + \langle 0|T(AD)|0\rangle  \langle 0|T(BC)|0\rangle 
 \end{eqnarray*}

 So,
 \begin{eqnarray*}
 i(\Delta_F)_{\mu\nu}^{ab}(x'-x) &=& (0|T\left[\chi^a_\mu(x') \chi^{b \dagger}_\nu(x)\right]|0)
 \nonumber\\ &\Leftarrow& 
 -\langle 0|T(AC)|0\rangle  \langle 0|T(BD)|0\rangle  + \langle 0|T(AD)|0\rangle  \langle 0|T(BC)|0\rangle 
 \end{eqnarray*}
 since in our case 
	$ \langle 0|T(AB)|0\rangle = \langle |T(CD)|0\rangle =0$. for the evaluation of these
	see Eqn.~(\ref{app:propa5e}).             
	\vspace{2mm}
% \end{minipage} }\\
% \end{center}

%%------------------------------------------------------------------------------
%% figuur 5
%\begin{center}
% \begin{figure}[hbt]
%\centering
%%\resizebox{7.25cm}{5.75cm}
%%\resizebox{7.25cm}{6.75cm}
% \resizebox{7.25cm}{6.75cm}
% {\includegraphics[180,495][400,675]{Fig.scalar-diquark-meson.ps}}
%%---------------------------------------------------------------------------------
%	\vspace{-2cm}
%	\caption{Scalar Diquark-exchange and Meson-exchange.  
%	Panel (a): diquark exchange. Panel (b): antiquark-quark exchange.}
%  \label{fig:trans.10}
%\end{figure}
%\end{center}
%%---------------------------------------------------------------------------------

%--------------------------------------------------------------------------
 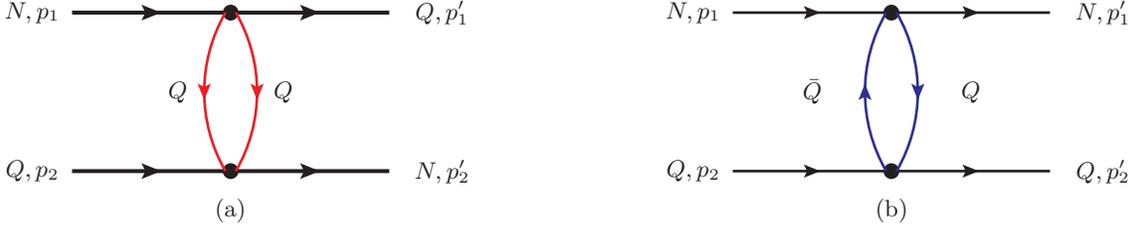
\begin{figure}[hbtp!]
%--------------------------------------------------------------------------
% axodraw figure:     
%\begin{center} \begin{picture}(350,175)(0,75)
%\begin{center} \begin{picture}(200,175)(0,75)
 \begin{center} \begin{picture}(200,175)(0,15)
 \SetPFont{Helvetica}{9}
 \SetScale{1.0} \SetWidth{1.5}
 \SetOffset(0,0)
 \SetOffset(-110,0)
 \ArrowLine(15,150)(75,150)   
 \ArrowLine(75,150)(135,150)   
 \Vertex(75,150){3}
 \Vertex(75, 90){3}
%\DashLine(75,90)(75,150){2}
%\Photon(75,150)(75,122){1}{3}
%\Text( 75,120)[]{$\bigotimes$}
%\Photon(75,90)(75,118){1}{3}
 \ArrowLine(15,90)(75,90)   
 \ArrowLine(75,90)(135,90)   
%---------------------------------------------------------------------------
 \SetWidth{0.3}
 \SetWidth{1.0}
%\SetColor{Red}
%\CArc(75,120)(30,-90,90)
 \SetColor{Red}
%\CArc( 25,120)(60,-30,30)  
 \ArrowArcn( 25,120)(60,30,-30)  
%\CArc(125,120)(60,150,210)
 \ArrowArc(125,120)(60,150,210)
 \SetColor{Black}
% \SetWidth{1.5}
%\DashArrowArc(75,120)(30,-90,90){1}
%\DashArrowArcn(75,120)(30,-90,90){1}
%---------------------------------------------------------------------------
 \Text( 0, 150)[]{$N,p_1$}
 \Text(155,150)[]{$Q,p_1'$}
 \Text( 0, 90)[]{$Q,p_2$}
 \Text(155,90)[]{$N,p_2'$}
 \Text(55,120)[]{$Q$}
 \Text(95,120)[]{$Q$}
 \Text(75, 75)[]{(a)}
%---------------------------------------------------------------------------
  \SetOffset(140,0)
  \ArrowLine(15,150)(75,150)   
  \ArrowLine(75,150)(135,150)   
  \Vertex(75,150){3}
  \Vertex(75, 90){3}
% \DashLine(75,90)(75,150){2}
% \Photon(75,150)(75,122){1}{3}
% \Text( 75,120)[]{$\bigotimes$}
% \Photon(75,90)(75,118){1}{3}
  \ArrowLine(15,90)(75,90)   
  \ArrowLine(75,90)(135,90)   
%---------------------------------------------------------------------------
 \SetWidth{0.5}
 \SetWidth{1.0}
%\SetColor{Red}
%\CArc(75,120)(30,-90,90)
 \SetColor{Blue}
%CArc( 25,120)(60,-30,30)  
 \ArrowArcn( 25,120)(60,30,-30)  
%\ArrowArc(125,120)(60,150,210)
 \ArrowArcn(125,120)(60,210,150)
 \SetColor{Black}
 \SetWidth{1.5}
%\DashArrowArc(75,120)(30,-90,90){1}
%\DashArrowArcn(75,120)(30,-90,90){1}
%---------------------------------------------------------------------------
 \SetWidth{1.5}
 \Text( 0, 150)[]{$N,p_1$}
 \Text(155,150)[]{$N,p_1'$}
 \Text( 0, 90)[]{$Q,p_2$}
 \Text(155,90)[]{$Q,p_2'$}
 \Text(45,120)[]{$\bar{Q}$}
 \Text(105,120)[]{$Q$}
 \Text(75, 75)[]{(b)}
%---------------------------------------------------------------------------
 \end{picture} \end{center}
\vspace{-2cm}
\caption{Scalar Diquark-exchange and Meson-exchange.  
Panel (a): diquark exchange. Panel (b): antiquark-quark exchange.}
 \label{fig:trans.10}
\end{figure}
%---------------------------------------------------------------------------------
%---------------------------------------------------------------------------------
 \section{Comparison Scalar Diquark-exchange and Meson-exchange}      
 \label{app:scalar-exchange}
%--------------------------------------------------------------------------
%{\red NOGAFTERONDEN}\\
 For meson-exchange we have an antiquark-quark state
 \begin{eqnarray*}
	 \psi^a(x) &=& \bar{q}^c(x) q^d(x) \epsilon^{acd}\ ,\ 
	 \bar{\psi}^a(x) = \bar{q}^d(x) q^c(x)\ \epsilon^{acd} 
	 =-\bar{q}^e(x) q^f(x) \epsilon^{aef}. 
 \end{eqnarray*}
 Then for the propagator
 \begin{eqnarray*}
	 &&	 (0|T\left[\psi^a(x') \bar{\psi}^b(x)\right]|0) = 
	 -\epsilon^{acd}\epsilon^{bef}\
	 (0|T\left[ \bar{q}^c(x') q^d(x')\cdot \bar{q}^e(x)\ q^f(x)\right]|0)
	 \rightarrow \nonumber\\ &&
	 +\epsilon^{acd}\epsilon^{bef}\
	 (0|T\left[q^d(x')\bar{q}^e(x)\right]|0)
	 (0|T\left[q^f(x)\bar{q}^c(x')\right]|0)
 \end{eqnarray*}
 The color factor becomes
	 $+\epsilon^{acd}\epsilon^{bef} \delta_{de}\delta_{cf} = -2\delta_{ab}$, 
 So, we get 
 \begin{eqnarray*}
	 (0|T\left[\psi^a(x') \bar{\psi}^b(x)\right]|0) = 
	 -2\delta_{ab}                 
	 (0|T\left[q(x')\bar{q}(x)\right]|0)
	 (0|T\left[q(x)\bar{q}(x')\right]|0)
 \end{eqnarray*}
 which differs indeed a (-)-sign with the corresponding (\ref{app:propa4e}).
 Since the rest of the calculations are identical this results in an overall (-)-sign.\\
 The difference between a meson and the diquark is the assumed absence of a bound state.
 In general, see {\it e.g.} \cite{NO90},
 \begin{eqnarray*}
	 \rho(s) &=& Z \delta(s-m_R^2) + \sigma(s)\ \theta(s-4m_R^2)
 \end{eqnarray*}
 with $Z \geq 0$ and $\sigma(s) \geq 0$ and
 \begin{eqnarray*}
	 1- Z &=& \int_{4m_R^2}^\infty ds\ \sigma(s)\ ,\ 0 \leq Z \leq 1.
 \end{eqnarray*}
 Because of the positive signs of both terms the presence or absence of a bound state 
 for a vector, axial-vector, etc mesons  makes no difference.\\
 {\bf Conclusion: 
 For the antiquark-quark exchange we get an extra (-)-sign 
 compared to the diquark-exchange. } 

\begin{figure}[htbp!]
%\begin{center} \begin{picture}(275,110)(0,0)
\begin{center} \begin{picture}(275,140)(0,30)
%\SetScale{1.25}
%\SetOffset(00,-140)
\SetWidth{1.2}
\SetColor{Red} 
\ArrowLine(145,110)(145, 10)
\ArrowLine(155,110)(155, 10)
\ArrowLine(20,10 )(150,10 )
 \CCirc(150,110){10}{Black}{Blue}
%\CCirc(150,110){10}{Black}{White}
%\Text(150,110)[c]{$\bigotimes$} 
\Text(150,130)[c]{$\langle q\Gamma q \rangle$}
\SetColor{Black}
\Line(150,  5)(250,  5)
\Line(150,10 )(250,10 )
\Line(150, 15)(250, 15)
\SetColor{Black}
%\CCirc(150,100){10}{Black}{Blue}
\CCirc(150,10 ){10}{Black}{PineGreen}

%\SetColor{PineGreen}
\Text(15,10 )[r]{$a,q_1$} \Text(135, 70)[r]{$b,q_2$} \Text(160, 70)[l]{$c,q_3$}
\Text(255,10 )[l]{N, p} 

\end{picture} \end{center}
	\vspace{8mm}
 \caption{Diquark-Quark-Nucleon vertex, Di-quark condensate 
$\big\langle \widetilde{q}^b \Gamma q^c \varepsilon^{abc} \big\rangle$
with $\Gamma= C \gamma_5$ from Cooper-pairing.}                             
\label{fig.diquark-quark-nucleon}                   
\end{figure}
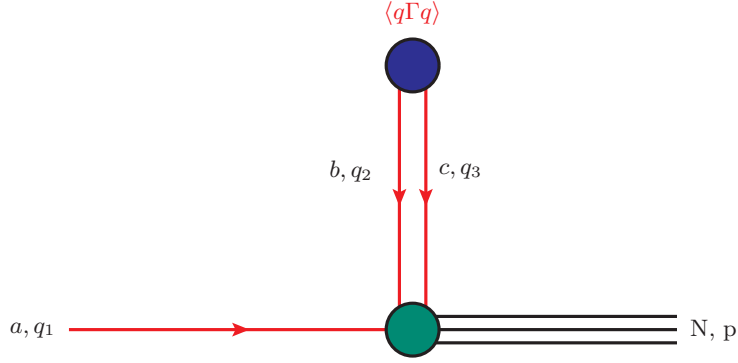                   
%-----------------------------------------------------------------------
%------------------------------------------------------------------------------
 \section{Scalar and Pseudoscalar Diquarks} \label{app:scal} 
%------------------------------------------------------------------------------
%{\blue Question: How can such a scalar Diquark give a $N \rightarrow Q$ 
%transition?? It seems implossible!? }\\
%------------------------------------------------------------------------------
\noindent In \cite{ARW98} a new form of ordering at high density, with color and
 flavor degrees of freedom correlated, where a condensate of scalar
 diquarks occurs. Therefore, we work out the exchange of such diquarks
 and introduce the isoscalar scalar field
 \footnote{Since $\gamma_0 C \gamma_0= -C$ the $\chi^a(x)$ is a scalar field.}
%------------------------------------------------------------------------------
 In \cite{Ioffe83} three representations of the proton as a three-quark system are
 discussed. For the representation with the axial-vector $\chi^a_\mu(x)$, we used 
 the $\eta^{(1)}$-current. Here we analyze the second current
 \begin{eqnarray}
 \eta^{(2)} &=& (\widetilde{u}^a C \sigma_{\mu\nu} u^b) \sigma^{\mu\nu}\gamma_5 d^c
 \varepsilon^{abc} = -4\left[(\widetilde{u}^a_R C d^b_R) u^c_R - 
 (\widetilde{u}^a_L C d^b_L) u^c_L\right]\ \varepsilon^{abc} 
 \nonumber\\ &=& -2\left[\left(\widetilde{u}^a C\gamma_5 d^b\right) u^c
 +\left(\widetilde{u}^a C d^b\right) \gamma_5 u^c\right] \varepsilon^{abc} 
 \equiv -2\left[ \chi_S^a u^a + \chi_5^a u^a\right],
 \label{app:scal.1a}\end{eqnarray}
 Note that in the first step $u \leftrightarrow d$ interchange has been performed
 using a Fierz-transformation in Dirac and isospin space. This current contains the
 composite diquark combinations 
 \begin{eqnarray}
 \chi_S^a = (\widetilde{u}^b C \gamma_5 d^c) \varepsilon^{abc},\ 
 \chi_5^a = (\widetilde{u}^b C d^c) \varepsilon^{abc},\ 
 \label{app:scal.1b}\end{eqnarray}
 where $\chi_S^a$ has spin-parity $J^{PC}=0^{++}$ and $\chi_5^a$ has $J^{PC}=0^{-+}$,
 Furthermore, $ud = [ |1,1/2)+|0,1,2)]/\sqrt{2}$ {\it i.e.} a combination of I=0,1.
 In quark matter a condensate with $\langle 0|\chi_S| 0\rangle \ne 0$ is possible, see
 \cite{ARW98} and the illustration in Fig.~\ref{app:scal.1c}.\\
\noindent The interaction Lagrangian for the NDQ-transition is
 \begin{eqnarray}
 {\cal L}_I^{(2)} &=& -\lambda_2\bigl\{\bar{\psi}\eta^{(2)}+ \bar{\eta}^{(2)}\psi\bigr\}
	 \nonumber\\ &=& 2\lambda_2 
	 \left\{\bar{\psi}\left[ \chi_S^a(x) u^a(x) + \chi_5^a(x) u^a(x)\right] + h.c.
	 \right\}.
 \label{app:scal.1c}\end{eqnarray}
%------------------------------------------------------------------------------
%{\blue Question: If $\chi_5$ is a scalar field, how does it couple to the
%quark and nucleon? With 1 or with gamma5? Does it give a scalar or a pseudo
%scalar potential fo Q +N $\rightarrow$ N+Q? 
%Because $\gamma_0 C \gamma_0 = -C$ $\chi_5$ is a scalar field!! So: revise this section!
%}
%------------------------------------------------------------------------------
 \begin{eqnarray}\label{app:scal.1}
 \chi_S^a(x) &=& \widetilde{q}^c(x) C\gamma_5 q^d(x)\ \varepsilon^{acd}\ ,\
 \chi_S^{b\dagger}(x) = -\bar{q}^e(x)\gamma_5C\widetilde{\bar{q}}^f(x)\
 \varepsilon^{bef}.
 \end{eqnarray}
 Following the same steps as for the $\chi^a_\mu(x)$ propagator, 
 analogous to (\ref{app:propa5h}) we have \\
\begin{eqnarray}
 X_S &=& 
4\delta_{ab} Tr\left[\gamma_5 \left(i\gamma\cdot\partial^y+m_Q\right) \gamma_5
\left(i\gamma\cdot\partial^z-m_Q\right)\right]\ \Delta_F(y)\cdot\Delta_F(z)   
\nonumber\\ &=& 
4\delta_{ab} \Tr\left[\gamma_\alpha\gamma_\beta \partial^y_\alpha\partial^z_\beta
- m_Q^2 \right]\ \Delta_F(y)\cdot\Delta_F(z) \nonumber\\ 
  &=& -16\delta_{ab} \left[ -\partial^y_\alpha \partial_z^\alpha 
	+m_Q^2 \right]\ \Delta_F(y)\cdot\Delta_F(z),   
 \label{app:scal.2a}\end{eqnarray}
and 
\begin{eqnarray}
 X_5 &=& 
4\delta_{ab} Tr\left[\left(i\gamma\cdot\partial^y+m_Q\right) 
\left(i\gamma\cdot\partial^z-m_Q\right)\right]\ \Delta_F(y)\cdot\Delta_F(z)   
\nonumber\\ &=& 
-4\delta_{ab} \Tr\left[\gamma_\alpha\gamma_\beta \partial^y_\alpha\partial^z_\beta
+ m_Q^2 \right]\ \Delta_F(y)\cdot\Delta_F(z) \nonumber\\ 
  &=& -16\delta_{ab} \left[ \partial^y_\alpha \partial_z^\alpha 
	+m_Q^2 \right]\ \Delta_F(y)\cdot\Delta_F(z),   
 \label{app:scal.2b}\end{eqnarray}
 where the differentiation variables, which in the end will be put to $y=z=x'-x$.\\
%------------------------------------------------------------------------------
 \subsection{Scalar Diquarks} \label{app:scal.a} 
%------------------------------------------------------------------------------
 The spectral representation of the Diquark Feynman propagator is 
 \begin{eqnarray}
	 i(\Delta_F)_{S,ab}(x'-x) &=& 
	 (0|T\left[\chi^a(x') \chi^{b \dagger}(x)\right]|0)
	 = i\int_{s_0}^\infty ds\ (\Delta_F)_{ab}(x'-x);s)\ \rho(s).
 \label{app:scal.3}\end{eqnarray}
 In momentum space the propagator leads to the integral
 \begin{eqnarray}
	 \widetilde{I}_{S}(k;m) &=& \int\frac{d^4p}{(2\pi)^4}\ 
	 \int\frac{d^4q}{(2\pi)^4}\ (2\pi)^4\delta^4(p+q-k)
	 \bigl[ p\cdot q+m^2\bigr]\ \times 
	 \left[p^2-m^2+i\epsilon\right]^{-1} \left[q^2-m^2+i\epsilon\right]^{-1}.
 \label{app:scal.4}\end{eqnarray}
 Following same steps below Eqn.~(\ref{app:disp.6}) leads now to 
 \begin{eqnarray}
 \widetilde{I}_S(k) &=& -i \sum_i \frac{c_i}{16\pi^2}
  \int_0^\infty \int_0^\infty \frac{dz_1dz_2}{(z_1+z_2)^2} 
 \left[ -\frac{2i}{(z_1+z_2)}+\frac{z_1z_2 k^2}{(z_1+z_2)^2}+m_i^2\right]
 \cdot\nonumber\\ && \times
 \exp\left\{i\left[k^2\frac{z_1z_2}{z_1+z_2}-(m_i^2-i\epsilon)(z_1+z_2)
 \right]\right\}.
 \label{app:scal.5}\end{eqnarray}
 %-----------------------------------------------------------------------------
 Now, see \cite{BD65}, Eqn.~(8.17), 
 \begin{eqnarray}
	  \widetilde{J}(k) &=& 
   \int_0^\infty \int_0^\infty \frac{dz_1dz_2}{(z_1+z_2)^2} 
	 \left[m^2 -\frac{i}{(z_1+z_2)}-\frac{k^2 z_1z_2}{(z_1+z_2)^2}\right]
 \cdot\nonumber\\ && \times
 \exp\left\{i\left[k^2\frac{z_1z_2}{z_1+z_2}-(m_i^2-i\epsilon)(z_1+z_2)\right]\right\}
 =0.
 \label{app:scal.6a}\end{eqnarray}
 So, under the integral the linear combination 
 \begin{eqnarray}
	 0 \equiv m^2 -\frac{i}{(z_1+z_2)}-\frac{k^2 z_1z_2}{(z_1+z_2)^2}        
 \label{app:scal.6b}\end{eqnarray}
 and we have three equivalent expressions for $\widetilde{I}_5(k)$:
 \begin{subequations}\label{app:scal.7}
 \begin{eqnarray}
	 (i)\ \widetilde{I}_S(k) &=& -i \sum_i \frac{c_i}{16\pi^2}
   \int_0^\infty \int_0^\infty \frac{dz_1dz_2}{(z_1+z_2)^2} 
  \left[2m_i^2 -\frac{3i}{(z_1+z_2)}\right]
 \cdot\nonumber\\ && \times
 \exp\left\{i\left[k^2\frac{z_1z_2}{z_1+z_2}-(m_i^2-i\epsilon)(z_1+z_2)\right]\right\} \\
	 (ii)\ \widetilde{I}_S(k) &=& -i \sum_i \frac{c_i}{16\pi^2}
   \int_0^\infty \int_0^\infty \frac{dz_1dz_2}{(z_1+z_2)^2} 
	 \left[ -\frac{i}{(z_1+z_2)}+2k^2\frac{z_1z_2}{(z_1+z_2)^2}\right]
 \cdot\nonumber\\ && \times
 \exp\left\{i\left[k^2\frac{z_1z_2}{z_1+z_2}-(m_i^2-i\epsilon)(z_1+z_2)\right]\right\} \\
	 (iii)\ \widetilde{I}_S(k) &=& -i \sum_i \frac{c_i}{16\pi^2}
   \int_0^\infty \int_0^\infty \frac{dz_1dz_2}{(z_1+z_2)^2} 
  \left[3k^2\frac{z_1z_2}{(z_1+z_2)^2}-m_i^2\right]
 \cdot\nonumber\\ && \times
 \exp\left\{i\left[k^2\frac{z_1z_2}{z_1+z_2}-(m_i^2-i\epsilon)(z_1+z_2)\right]\right\}.  
 \end{eqnarray}\end{subequations}
 %-----------------------------------------------------------------------------
 \noindent Using the identity (\ref{app:disp.10b})              
 in $\widehat{I}_{\mu\nu}(k)$ and subsequently scaling $z_i \rightarrow \lambda z_i$,
 one gets for (iii) the expression
 \begin{eqnarray}
	 \widetilde{I}_S(k) &=&  \frac{-i}{16\pi^2}\ \sum_i c_i\ 
 \int_0^\infty \int_0^\infty dz_1 dz_2\ \delta(1-z_1-z_2)\ 
	 \int_0^\infty \frac{d\lambda}{\lambda}
 \left[3z_1z_2 k^2-m_i^2\right]
 \cdot\nonumber\\ && \times 
 \exp\left[i\lambda\left(z_1z_2 k^2-m_i^2+i\epsilon\right) \right].
 \label{app:scal.8}\end{eqnarray}
 The $\lambda$-integral diverges logarithmically and is evaluated by applying 
 the cut-off procedure by choosing $C_1=-1, C_i=0\ (i > 1)$. This gives
 \begin{eqnarray}
	 \widehat{I}_S(k) &=& 
	 \widetilde{I}_S(k; m^2) - \widetilde{I}_S(k; M^2) \nonumber\\ 
	 &\approx& -i(16\pi^2)^{-1}\ \sum_i c_i\ 
 \int_0^\infty \int_0^\infty dz_1 dz_2\ \delta(1-z_1-z_2)\ 
 \left[ 3z_1z_2 k^2-m_i^2\right]
 \cdot\nonumber\\ && \times  
	 \int_0^\infty \frac{d\lambda}{\lambda}
 \exp\left[i\lambda\left(z_1z_2 k^2-m_i^2+i\epsilon\right) \right] \\
 &=& +(16\pi^2)^{-1} \sum_{i=0}^1 c_i 
	 \int_0^1 dz\ \left[3z(1-z) k^2-m_i^2\right] 
	 \ln\left[1-z(z-1)\ k^2/m_i^2\right]
 \label{app:scal.9}\end{eqnarray}
 where in the last step we scaled $\lambda \rightarrow m_i^2\lambda$.
 Working this further out we get, with $m=m_Q$, 
 \begin{eqnarray}
 \widehat{I}_S(k) &=& +(16\pi^2)^{-1} \int_0^1 dz\ \biggl\{ 
 \left[3z(1-z) k^2\right]\ \ln\left[\frac{1-z(1-z) k^2/m^2}{1-z(1-z)k^2/M^2}\right]
 \nonumber\\ 
 &&  -\left[ m^2\ln\left\{1-z(1-z) k^2/m^2\right\}
 -M^2\ln\left\{1-z(1-z) k^2/M^2\right\} \right] \biggr\}
 \label{app:scal.18}\end{eqnarray}
 Using the approximation $\ln(1-a/m_i^2) \approx -a/m_i^2$ for
 $m_i=m,M$ we obtain
 \begin{subequations}\label{app:scal.19}
 \begin{eqnarray}
 && \widehat{I}_S(k) \approx +(3/16\pi^2) \int_0^1 dz\ 
 \left[z(1-z) k^2\right]\ \left(\ln\frac{M^2}{m^2}+ 
 \ln\left[\frac{m^2-z(1-z) k^2}{M^2-z(1-z)k^2}\right]\right)
 \nonumber\\ && \approx 
+(3/16\pi^2) \int_0^1 dz\  
 \left[z(1-z) k^2\right]\ \left(\ln\frac{M^2}{m^2}- 
 \ln\left[\frac{M^2}{m^2-z(1-z) k^2}\right]\right)
% \nonumber\\ 
% &&  +\left[ \left(m^2\ln m^2-M^2\ln\ M^2\right)
% +m^2\ln\left\{1-z(1-z) k^2/m^2\right\}
% -M^2\ln\left\{1-z(1-z) k^2/M^2\right\} \right] \biggr\}
% &&\approx  (16\pi^2)^{-1} \biggl\{+6k^2\ln\frac{M^2}{m^2}
%	 + \left(m^2\ln m^2-M^2\ln\ M^2\right) \nonumber\\ && 
%	 +m^2\int_0^1 dz\ \left[1-z(1-z)k^2/m^2\right] \ln\left[1-z(1-z)k^2/m^2\right]
	 \nonumber\\ &=&
	+(3/16\pi^2) k^2\biggl\{ 
	\frac{1}{6}\ln\frac{M^2}{m^2}-\int_0^1 dz\ z(1-z)\
	\ln\left[\frac{M^2}{m^2-z(1-z) k^2}\right]\biggr\}
 \end{eqnarray}\end{subequations}
	 Since for the nucleon-quark vertex $k^\mu \approx (m_N-m_Q, {\bf k})$ we
	 take $k^2 \cong (m_N-m_Q)^2$ in the front factor. 
% {\red tothier}\\
 The z-integral is 
 \begin{eqnarray*}
% I_1(k) &=& \int_0^1 dz\ \ln\left[1-z(1-z) k^2/m^2\right] =
%% \ln\left(\frac{k^2}{m^2}\right)+ \int_0^1 dz\ \ln[(z-z_+)(z-z_-)] = 
% %\ln\left(\frac{k^2}{m^2}\right) +\ln\left(\frac{m^2}{k^2}\right) +
% \sqrt{1-\frac{4m^2}{k^2}}\ln\left[\frac{1+\sqrt{1-4m^2/k^2}}{1-\sqrt{1-4m^2/k^2}}\right], \\
 I_2(k) &=& \int_0^1 dz\ z(1-z)\ln\left[1-z(1-z) k^2/m^2\right] =
  \frac{1}{6}\biggl\{ (1+2A) \sqrt{1-4A}\ 
 \ln\left[\frac{1+\sqrt{1-4A}}{\left|1-\sqrt{1-4A}\right|}\right]
 -\left(4A+\frac{5}{3}\right)\biggr\}.     
 \end{eqnarray*}
 where the $I_2$-integral has been taken from (\ref{app:disp.17}) and $A=m^2/k^2$.
 So, we finally get
 \begin{eqnarray}
	 \widehat{I}_S(k) &\approx& +(3/16\pi^2) k^2\ I_2(k)
 \label{app:scal.20}\end{eqnarray}
% The (unrenormalized) diquark propagator becomes
% \begin{eqnarray}
%	 i\left(\widetilde{\Delta}^{(0)}_F\right)_{S,ab}(k) &=& 
%	 -i\frac{3k^2}{\pi^2}\ 
%	 \times \left[\int_0^1 dz\ z(1-z)\ \ln\left(1-z(1-z)\
%	 \frac{k^2}{m^2}\right)\right]\ \delta_{ab}.
% \label{app:scal.21}\end{eqnarray}
%In the limit $k^2 \rightarrow 0$, the propagator is a constant $Z_3$, defined by
%\begin{eqnarray}
% Z_3 &=& 1-\frac{g_0^2}{3\pi} \ln\left(\frac{M^2}{m^2}\right), 
%\label{app:scal.22}\end{eqnarray}
%and the renormalized coupling is defined as
%\begin{eqnarray}
% g^2_R &=& Z_3 g_0^2 = 
%  g_0^2\left[1-\frac{g_0^2}{3\pi} \ln\left(\frac{M^2}{m^2}\right)\right], 
%\label{app:scal.23}\end{eqnarray}
%which is independent of the momentum transfer.

%\noindent {\it The $\chi^a$-field used thus far is not normalized to dimension [MeV]. We now normalize 
%	 by redefining $\chi^a(x) \rightarrow \chi^a(x) (\hbar c/m_\chi)^2$.}
%Furthermore, since in perturbation theory the free propagators are used we use in the projection
%operator $k^2=m_\chi^2$.}  
The  "unrenormalized" propagator becomes
 \begin{eqnarray}
	 i(\widetilde{\Delta}_F)_{S,ab}(k) &=& 
	 -i\frac{3k^2}{\pi^2}\ \times\left[\int_0^1 dz\ z(1-z)\ \ln\left(1-z(1-z)\
	 \frac{k^2}{m^2}\right)\right]\ \delta_{ab}
 \label{app:scal.24}\end{eqnarray}
 which gives for the "renormalized propagator", {\it i.e.} the finite part
 \footnote{This dispersive method of "renormalization" is from \cite{Kallen52}, and is an
 alternative to the Pauli-Villars method.},
 \begin{eqnarray}
	 i (\widetilde{\Delta}_F)_{S,ab}(k) &=& 
	 -i\frac{3k^2}{12\pi^2} \int_{4m^2}^{\infty} ds\
	\left(1+\frac{2m^2}{s}\right)\sqrt{1-\frac{4m^2}{s}} 
	\bigl[s\left(k^2-s+i\epsilon\right)\bigr]^{-1} \delta_{ab}.
 \label{app:scal.25}\end{eqnarray}
 In the process $N \rightarrow Q+D$ we approximate in the numerator 
 $k^2 \approx k_0^2 \approx (m_N-m_Q)^2i=(\Delta m_{NQ})^2$   
 valid for low-momentum transfer.\\
 %-----------------------------------------------------------------------------------------
 The spectral representation of the Diquark Feynman propagator is 
 \begin{eqnarray}
	 i(\Delta_F)_{S,ab}(x'-x) &=& 
	 (0|T\left[\chi^a(x') \chi^{b \dagger}(x)\right]|0)
	 = i\int_{4m^2}^\infty ds\ (\Delta_F)_{S,ab}(x'-x;s)\ \rho(s)
	 \nonumber\\ &=& -i \delta_{ab}\ \left(\frac{m_N-m_Q}{\hbar c}\right)^2
	 D(x'-x;m_\chi,\Lambda).  
 \label{app:scal.26}\end{eqnarray}
 %-----------------------------------------------------------------------------------------
 The Fourier transforms, with $m_\chi \sim 2 m_Q$, are
 \begin{subequations}\label{app:scal.27}
 \begin{eqnarray}
	 \widetilde{D}(k^2,m_\chi^2) &=& -\left(\frac{\Delta m_{NQ}}{\hbar c}\right)^2
	 \int_{4m_Q^2}^{\infty} ds\ \rho(s)\ \widetilde{\Delta}_F(k^2,\Lambda^2;s), \\
 \widetilde{\Delta}_F(k^2;s) &=& \left[k^2-s+i\epsilon \right]^{-1},\ \
 \rho(s) = \frac{1}{12\pi^2} \left(1+\frac{2m^2}{s}\right) \sqrt{1-\frac{4m^2}{s}}/s.
 \end{eqnarray}\end{subequations}
	 With $\sqrt{s} = E({\bf q})= \sqrt{{\bf q}^2+m_\chi^2}$ 
	 we have  $k^2\approx \Delta m_{NQ}^2-{\bf k}^2$ and
 \begin{subequations}\label{app:scal.28}
 \begin{eqnarray}
%\widetilde{\Delta}_F({\bf k}^2,\Lambda^2;s) &=&+\left[{\bf k}^2+{\bf q}^2+m_\chi^2\right]^{-1}, \ \
	 \widetilde{D}({\bf k}^2,m_\chi^2) &=& 4\pi \left(\frac{\Delta m_{NQ}}{\hbar c}\right)^2
  \int_{0}^\infty q^2dq\ \rho(s)\ \widetilde{\Delta}_F({\bf k}^2,\Lambda^2;s), \\    
	 \widetilde{\Delta}_F({\bf k}^2,\Lambda^2;s) &=& \exp\left[-{\bf k}^2/\Lambda^2\right]
	 \left[{\bf k}^2+{\bf q}^2+m_\chi^2\right]^{-1}, 
%\label{app:disp.33} \end{eqnarray}
 \end{eqnarray}\end{subequations}
 where we added a gaussian form-factor as usual in the Nijmegen potentials.
%-----------------------------------------------------------------------------------------
 In configuration space
 \begin{eqnarray}
 D({\bf x}^2,m_Q^2) &=& 
	 \left(\frac{\Delta m_{NQ}}{\hbar c}\right)^2 \int_{4m_Q^2}^{s_{max}} ds\ \rho(s)
 \left[\frac{m(s)}{4\pi}\ \phi_C^0\left(m(s),\Lambda; |{\bf x}|\right)\right]
 \label{app:scal.29}\end{eqnarray}
 where $m(s)= \sqrt{s-m_\chi^2}$.

%%--not relevant: --------------------------------------------------------------
%\noindent  The pseudoscalar diquark-exchange, see Fig.~\ref{fig:trans.1a}, gives the 
% QN-potential, see {\it e.g.} \cite{RNY2010},
% \begin{eqnarray}
% V_{QN}(r) &=& -\lambda_5^2 \frac{m_\chi^2}{4\pi}
% \left[\frac{m_\chi^2}{4m_Q m_N}\left(\frac{1}{3} (\bm{\sigma}_1\cdot\bm{\sigma}_2)
%	 \phi^1_C(r)+ S_{12} \phi_T^0\right)\right]\ P_x.
% \label{app:scal.31}\end{eqnarray}
% {\it We notice that the volume integral of this potential is zero, which is 
% different from that from $\chi_\mu^a$-exchange. Therefore, it is expected to be
% weaker.}

%------------------------------------------------------------------------------
 \noindent Approximating the propagator with an effective one for $k^2 <0$, 
 \begin{eqnarray}
 (\widetilde{\Delta}(k^2))_{S,ab} &\cong & -\left(\frac{\Delta_{NQ}}{\hbar c}\right)^2
 \frac{F(k^2)}{k^2-\bar{m}_{\chi}^2+i\epsilon} \approx
 +\left(\frac{\Delta_{NQ}}{\hbar c}\right)^2
 \frac{\exp(-{\bf k}^2/\Lambda^2)}{{\bf k}^2+\bar{m}_\chi^2}, 
 \label{app:scal.30}\end{eqnarray}
	 where $\bar{m}_\chi^2= m_\chi^2-\Delta_{NQ}^2 >0$,
the scalar diquark-exchange, see Fig.~\ref{fig:trans.1a}, gives the 
 QN-potential, see {\it e.g.} \cite{RNY2010},
 \begin{eqnarray}
	 V_{QN}(r) &=& +\bar{\lambda}_S^2 \frac{\bar{m}_\chi}{4\pi}
	 \biggl\{\left[\phi_C^0-\frac{\bar{m}_\chi^2}{4m_N m_Q} \phi_C^1\right]
	 +\frac{\bar{m}_\chi^2}{2m_N m_Q} \phi_{SO}^0 {\bf L}\cdot{\bf S}
	 +\frac{\bar{m}_\chi^4}{16m_N^2 m_Q^2}\cdot\nonumber\\ 
	 && \times \frac{3}{(\bar{m}_\chi r)^2} \phi_T^0 Q_{12}
	 +\frac{m_Q^2}{m_N m_Q}\left[\frac{(m_N^2-m_Q^2)}{4m_N m_Q}\right]\ \phi_{SO}^0
 \cdot\frac{1}{2}(\bm{\sigma}_1-\bm{\sigma}_2)\cdot{\bf L}
	 \nonumber\\ && 
 +\frac{1}{4m_Nm_Q}\left(\bm{\nabla}^2\phi_C^0+\phi_C^0\bm{\nabla}^2\right)
 \biggr\}\ P_x. 
 \label{app:scal.31a}\end{eqnarray}
 where $\bar{\lambda}_S^2 (\Delta m_{NQ}/\hbar c)^2$, and 
 with a factor $\langle 1 + \bm{\tau}_1\cdot\bm{\tau}_2\rangle/2 = 2I-1$.
 For $NQ \rightarrow QN$ this gives for S-waves, averaging over the spin with factors
 1/4 and 3/4 for the singlet and triplet respectively, the spin-isospin factor is -1/2,
 which gives attraction.
%which is repulsive for L=0,2 4, .... and attractive for L=1,3,5, ...
%------------------------------------------------------------------------------
 \subsection{Pseudoscalar Diquarks} \label{app:scal.b} 
%------------------------------------------------------------------------------
 The spectral representation of the Diquark Feynman propagator is 
 \begin{eqnarray}
	 i(\Delta_F)_{5,ab}(x'-x) &=& 
	 (0|T\left[\chi_5^a(x') \chi_5^{b \dagger}(x)\right]|0)
	 = i\int_{s_0}^\infty ds\ (\Delta_F)_{ab}(x'-x);s)\ \rho(s).
 \label{app:psscal.3}\end{eqnarray}
 In momentum space the propagator leads to the integral
 \begin{eqnarray}
	 \widetilde{I}_{5}(k;m) &=& \int\frac{d^4p}{(2\pi)^4}\ 
	 \int\frac{d^4q}{(2\pi)^4}\ (2\pi)^4\delta^4(p+q-k)
	 \bigl[-p\cdot q+m^2\bigr]\ \times 
	 \left[p^2-m^2+i\epsilon\right]^{-1} \left[q^2-m^2+i\epsilon\right]^{-1}.
 \label{app:psscal.4}\end{eqnarray}
 Following same steps below Eqn.~(\ref{app:disp.6}) leads now to 
 \begin{eqnarray}
 \widetilde{I}_5(k) &=& -i \sum_i \frac{c_i}{16\pi^2}
  \int_0^\infty \int_0^\infty \frac{dz_1dz_2}{(z_1+z_2)^2} 
 \left[ +\frac{2i}{(z_1+z_2)}-\frac{z_1z_2 k^2}{(z_1+z_2)^2}+m_i^2\right]
 \cdot\nonumber\\ && \times
 \exp\left\{i\left[k^2\frac{z_1z_2}{z_1+z_2}-(m_i^2-i\epsilon)(z_1+z_2)
 \right]\right\},
 \label{app:psscal.5}\end{eqnarray}
 which is the analogon of Eqn.~(\ref{app:scal.5}) for the scalar case.
 Similar to (\ref{app:scal.8}) we now get 
 \begin{eqnarray}
	 \widetilde{I}_5(k) &=&  \frac{+3i}{16\pi^2}\ \sum_i c_i\ 
 \int_0^\infty \int_0^\infty dz_1 dz_2\ \delta(1-z_1-z_2)\ 
	 \int_0^\infty \frac{d\lambda}{\lambda}
 \left[z_1z_2 k^2-m_i^2\right]
 \cdot\nonumber\\ && \times 
 \exp\left[i\lambda\left(z_1z_2 k^2-m_i^2+i\epsilon\right) \right].
 \label{app:psscal.6}\end{eqnarray}
 Analogous to the scalar case treated above, the finite part
 \begin{eqnarray}
	 i (\widetilde{\Delta}_F)_{5,ab}(k) &=& 
	 +\frac{3i}{12\pi^2} 
	 \frac{k^2}{(\hbar c)^2} \int_{4m^2}^{\infty} ds\
	\left(1+\frac{2m^2}{s}\right)\sqrt{1-\frac{4m^2}{s}} 
	\bigl[s\left(k^2-s+i\epsilon\right)\bigr]^{-1} \delta_{ab}.
 \label{app:psscal.7}\end{eqnarray}
 and we can follow the same steps (\ref{app:scal.26}) $\rightarrow$ (\ref{app:scal.29}) giving
 for the "effective" propagator
 \begin{eqnarray}
 (\widetilde{\Delta}(k^2))_{5,ab} &\cong & +\left(\frac{\Delta_{NQ}}{\hbar c}\right)^2
 \frac{F(k^2)}{k^2-\bar{m}_{\chi}^2+i\epsilon} \approx
 -\left(\frac{\Delta_{NQ}}{m_\chi}\right)^2
 \frac{\exp(-{\bf k}^2/\Lambda^2)}{{\bf k}^2+\bar{m}_\chi^2}, 
 \label{app:psscal.8}\end{eqnarray}
 which is analogous to (\ref{app:scal.30}).

 %-----------------------------------------------------------------------------
\noindent  Then, the pseudoscalar diquark-exchange, see Fig.~\ref{fig:trans.1a}, gives the 
 QN-potential, see {\it e.g.} \cite{RNY2010},
 \begin{eqnarray}
	 V_{QN}(r) &=& -\bar{\lambda}_5^2 \frac{m_\chi^2}{4\pi}
 \left[\frac{m_\chi^2}{4m_Q m_N}\left(\frac{1}{3} (\bm{\sigma}_1\cdot\bm{\sigma}_2)
	 \phi^1_C(r)+ S_{12} \phi_T^0\right)\right]\ P_x, 
 \label{app:scal.31b}\end{eqnarray}
 where $\bar{\lambda}_5^2= \lambda_5^2 (\Delta m_{NQ}/\hbar c)^2$, 
 and with a factor $\langle 1 + \bm{\tau}_1\cdot\bm{\tau}_2\rangle/2 = 2I-1$.
 This gives, with $m_\chi=2m_Q$, for $NQ \rightarrow QN$ in $^1S_0$ a coupling 
 $+\lambda_5^2/3$ and in
 $^3S_1$ a coupling $+\lambda_5^2/3$. So, repulsion for S-waves.
% For $NQ \rightarrow QN$ this gives zero on average.
 {\it We notice that the volume integral of this potential is zero, which is 
 different from $\chi_\mu^a$-exchange. Therefore, it is expected to be
 weaker than the axial-vector and scalar diquark-exchange potential.}
 Note that from (\ref{app:scal.1c}) the scalar and pseudoscalar couplings are equal: 
 $\lambda_S=\lambda_5=2\lambda_2$.\\
\begin{flushleft}
\rule{16cm}{0.5mm}
\end{flushleft}
%------------------------------------------------------------------------------
%------------------------------------------------------------------------------
% Not in this version for nn-online! 
%%------------------------------------------------------------------------------
%% figuur 10
%%\vspace*{-1.5cm}
% \begin{center}
%  \begin{figure}[hbtp!]
% \centering
% %{\includegraphics*[width=16cm,height=21cm]{./Fig.triquark-nucleon.ps}}
%% \resizebox{7.25cm}{5.75cm}
%% \resizebox{7.25cm}{!}
% \resizebox{6.25cm}{3.25cm}
%  {\includegraphics[180,475][400,655]{Fig.QandN.graphs.ps}}
%% {\includegraphics[180,575][400,655]{Fig.QandN.graphs.ps}}
%  \vspace{+3.0cm}
%	  \caption{
%  	Panel (a): $B^\dagger B$-term NQ $\rightarrow$ QN,
%	Panel (b): $A^\dagger A$-term QQ $\rightarrow$ QQ,
%	Panel (c):  DQ $\rightarrow$ N, Panel (d): N $\rightarrow$ DQ.  }
% 	\label{fig:QandN-graph}       
%  \end{figure} \end{center}
%%---------------------------------------------------------------------------------
%-------------------------------------------------------------------------
% Contact-graphs: NQQN, QQQQ, NDQ, DQN
%-------------------------------------------------------------------------
% axodraw in Tex:
%-------------------------------------------------------------------------
\begin{figure}[htbp!]
%\begin{center} \begin{picture}(275,110)(0,0)
\begin{center} \begin{picture}(575,340)(0,30)
%\SetScale{1.25}
\SetScale{1.0}   
\SetOffset(-100,0)
%\SetOffset(-100,200)
\SetWidth{1.2}
\SetColor{Red} 
\Line(20,40)(150,100)
%\Line(20,100)(150,100)
%\Line(20,160)(150,100)
\SetColor{Blue}
\Photon(20,160)(150,100){3}{6.5}
\SetColor{Black}
	\Line(150, 95)(250, 95)
	\Line(150,100)(250,100)
	\Line(150,105)(250,105)
\SetColor{Black}
%\CCirc(150,100){10}{Black}{Blue}
\CCirc(150,100){10}{Black}{PineGreen}
%\SetColor{PineGreen}
\Text(15,160)[r]{$D^a$} \Text(15,40)[r]{$Q^a$}
\Text(255,100)[l]{N} 
\Text(150,30)[]{(c)}
%-------------------------------------------------------------------------
%\SetOffset(180,0)
\SetOffset(160,0)
\SetWidth{1.2}
\SetColor{Red} 
\Line(150,100)(280,40) 
\SetColor{Blue}
\Photon(150,100)(280,160){3}{6.5}
\SetColor{Black}
\Line( 20, 95)(150, 95)
\Line( 20,100)(150,100)
\Line( 20,105)(150,105)
\SetColor{Black}
%\CCirc(150,100){10}{Black}{Blue}
\CCirc(150,100){10}{Black}{PineGreen}
%\SetColor{PineGreen}
\Text(300,160)[r]{$D^a$} \Text(300,40)[r]{$Q^a$}
\Text(10,100)[l]{N} 
\Text(150,30)[]{(d)}
%-------------------------------------------------------------------------
%\end{picture} \end{center}
%%\caption{\sl Diquark-quark-Nucleon Vertex.}
%%\label{fig.triquark-nucleon}\end{figure}                   
%\end{figure}                   
%%-----------------------------------------------------------------------
%--------------------------------------------------------------------------
% Diquark-NQQN diagram
%--------------------------------------------------------------------------
%\begin{figure}[hbtp!]
%--------------------------------------------------------------------------
% axodraw figure:     
%\begin{center} \begin{picture}(350,175)(0,75)
%\begin{center} \begin{picture}(200,175)(0,75)
%\SetPFont{Helvetica}{9}
%\SetScale{1.0} \SetWidth{1.5}
%\SetOffset(110,110)
%\SetOffset(-100,0)
 \SetOffset(0,0)
 \ArrowLine(15,250)(75,220)   
 \ArrowLine(75,220)(135,250)   
 \ArrowLine(15,190)(75,220)   
 \ArrowLine(75,220)(135,190)   
 \Text( 75,220)[]{$\bigotimes$}
 \Text( 0, 250)[]{$N$}
 \Text(155,250)[]{$Q'$}
 \Text( 0,190)[]{$Q$}
 \Text(155,190)[]{$N'$}
\Text(75,180)[]{(a)}
\SetOffset(180,0)
 \ArrowLine(15,250)(75,220)   
 \ArrowLine(75,220)(135,250)   
 \ArrowLine(15,190)(75,220)   
 \ArrowLine(75,220)(135,190)   
 \Text( 75,220)[]{$\bigotimes$}
 \Text( 0, 250)[]{$Q$}
 \Text(155,250)[]{$Q'$}
 \Text( 0,190)[]{$Q$}
 \Text(155,190)[]{$Q'$}
\Text(75,180)[]{(b)}
%---------------------------------------------------------------------------
 \end{picture} \end{center}
 \caption{Diquark-exchange  for $NQ \rightarrow QN$ reaction.}                         
  \vspace{+3.0cm}
\caption{
  Panel (a): $B^\dagger B$-term NQ $\rightarrow$ QN,
Panel (b): $A^\dagger A$-term QQ $\rightarrow$ QQ,
Panel (c):  DQ $\rightarrow$ N, Panel (d): N $\rightarrow$ DQ.  }
 \label{fig:QandN-graph}       
  \end{figure}
%---------------------------------------------------------------------------------

%------------------------------------------------------------------------------
 \section{ DQE Interaction and the Partition Functional }  
 \label{app:DQEXCH-PF}
%%-------------------------------------------------------------------------------
%--------------------------------------------------------------------------
%\begin{center}
% \fbox{ \begin{minipage}[b]{16cm}
%\vspace{2mm}
%--------------------------------------------------------------------------
%\begin{center}
% {\blue \underline{Intermezzo}:}
%\end{center}
%	 \vspace{-5mm}
%--------------------------------------------------------------------------
  In this Appendix  
% \footnote{This is a correction of the Appendix in \cite{YYR24}. We stress that
% the potential $V^{(qN)}_{DQB}(r)$ is correct as well as the results in \cite{YYR24}.}
 the diquark exchange potential $V_{DQE}^{(qN)}$ is derived from a description 
 of the confinement-deconfinement process at high baryon-densities via the 
 (density dependent) nucleon-triquark coupling 
 \begin{eqnarray}
 	{\cal L}^{(1)}_{int} = -\lambda_3 \bigl[\bar{\psi}(x) \eta_N(x) 
 	+ \bar{\eta}_N(x) \psi(x)\bigr] 
 \label{eq:Ldeconf1a} \end{eqnarray}
 	with \cite{Ioffe81}
 \begin{eqnarray}
 \eta_N(x) = \left[\widetilde{q}^a(x)C\gamma^\mu q^b(x)\right] \gamma_5\gamma_\mu q^c(x) 
 	\varepsilon^{abc},
 \end{eqnarray}
 where C is the charge conjugation operator, and momentarily we left out the isospin labels.\\
 The interaction Lagrangian in (\ref{eq:Ldeconf1a}) with the tri-quark field $\eta_N(x)$ 
 is rewritten using di-quark fields for two reasons:  
 (i) the functional form of the partition function is difficult to handle,    
 and (ii) di-quarks are meaningful physical entities.
 In terms of the (bosonic) di-quark fields $\chi_\mu^a(x)$ 
 \begin{subequations}
 \begin{eqnarray}
 \eta_N(x) &=& (\hbar c)^2 \gamma_5\gamma^\mu q^a(x)\cdot \chi_\mu^a(x), \\
 \chi_\mu^a(x) &\equiv& \varepsilon^{abc} \widetilde{q}^b(x)C\gamma_\mu q^c(x)/(\hbar c)^2.
 \end{eqnarray}
 \end{subequations}
%--------------------------------------------------------------------------
% The interaction (\ref{eq:Ldeconf1a}) becomes
% \begin{eqnarray}
% 	{\cal L}^{(1)}_{int}  = -\lambda_3 (\hbar c)^2 \bigl[
% \left(\bar{\psi}(x)\gamma_5\gamma^\mu q^a(x)\right) \chi^a_\mu(x) + h.c. \bigr]  
% \label{eq:Ldeconf2a} \end{eqnarray}
% \begin{subequations}
% \begin{eqnarray}
% \eta_N(x) &=& (\hbar c)^2 \gamma_5\gamma^\mu q^a(x)\cdot \chi_\mu^a(x), \\
% \chi_\mu^a(x) &\equiv& \varepsilon^{abc} \widetilde{q}^b(x)C\gamma_\mu q^c(x)/(\hbar c)^2.
% \end{eqnarray}
% \end{subequations}
%--------------------------------------------------------------------------
 Here $\chi^a_\mu(x)$ is an auxiliary field a la \cite{Bender77}.
The interaction (\ref{eq:Ldeconf1a}), with $\bar{\lambda}_3 \equiv \lambda_3 (\hbar c)^2$, becomes
\begin{eqnarray}
	{\cal L}^{(1)}_{int}  = -\bar{\lambda}_3 \bigl[
\left(\bar{\psi}(x)\gamma_5\gamma^\mu q^a(x)\right) \chi^a_\mu(x) + h.c. \bigr]  
\label{eq:Ldeconf2a} \end{eqnarray}
and the auxiliary field Lagrangian 
\begin{equation}
	{\cal L}^{(0)}_\chi = \biggl\{\chi^{a \dagger}_\mu(x) \chi^{\mu a}(x) 
	- \bigl[\chi_\mu^{a \dagger}(x) 
	\bigl(\tilde{q}^b(x)C\gamma^\mu q^c(x)\bigr) \varepsilon^{abc} + {\it h.c.}\bigr]\biggr\}
\end{equation}
This gives for the Lagrangian including the interaction (\ref{eq:Ldeconf2a}) 
\begin{eqnarray}
	{\cal L}(\chi, q, \psi) &=&\biggl\{\chi^{a \dagger}_\mu(x) \chi^{\mu a}(x) 
	-\bigl[\chi_\mu^{a \dagger}(x) 
	\bigl(\tilde{q}^b(x)C\gamma^\mu q^c(x)\bigr) \varepsilon^{abc} + {\it h.c.}\bigr]\biggr\}
	\nonumber\\ && 
	 -\bar{\lambda}_3 \bigl[
\left(\bar{\psi}(x)\gamma_5\gamma^\mu q^a(x)\right) \chi^a_\mu(x) + h.c. \bigr]  
\label{eq:Ldeconf2b} \end{eqnarray}
Writing 
\begin{eqnarray*}
A^a_\mu  &\equiv& \varepsilon^{abc} \widetilde{q}^b(x)C\gamma_\mu q^c(x)/(\hbar c)^2\ ,\ 
	B^a_\mu \equiv  \bigl[ \left(\bar{\psi}(x)\gamma_5\gamma^\mu q^a(x)\right)   
\end{eqnarray*}
the Lagrangian becomes
\begin{eqnarray}
{\cal L}(\chi, q, \psi) &=&\biggl\{\chi^{a \dagger}_\mu(x) \chi^{\mu a}(x) 
-\bigl[\chi_\mu^{a \dagger}(x) A^a_\mu(x) + A_\mu^{a \dagger}(x) \chi^{a \mu}(x) \biggr\}
\nonumber\\ && 
 -\bar{\lambda}_3 \bigl[
	 B^{a \dagger}_ \mu(x)\chi^{a \mu}(x) + \chi^{a \dagger}_\mu(x) B^{a \mu}(x) \bigr]  
\label{eq:Ldeconf2c} \end{eqnarray}
Then, the EL equation $ \partial{\cal L}_\chi/\partial\chi^a=0$, gives
\begin{eqnarray*}
\chi^a_\mu &=& A_\mu^a +\bar{\lambda}_3 B_\mu^a.
%\approx A_\mu^a. 
\end{eqnarray*}
%\begin{eqnarray*}
%	{\cal L}_\chi &=& -\bar{\lambda}_3^2 B_\mu^{a \dagger} B^{a \mu}
%\end{eqnarray*}
%which is of first order in the deconfinement coupling.
		 Substituting this into ${\cal L}(\chi,q,\psi)$ gives the mean-field (MF) approximation
\begin{eqnarray}
	{\cal L}(\chi,q,\psi) &=& -{\it Sign}\ \biggl\{A^{a \dagger}_\mu A^{\mu a}+
	 \bar{\lambda}_3\left(A^{a \dagger}_\mu B^{\mu a}+
	B^{a \dagger}_\mu A^{\mu a}\right)
	+\bar{\lambda}_3^2 B^{a \dagger}_\mu B^{\mu a}
	\biggr\}.
\label{eq:Ldeconf2d} \end{eqnarray}
{\it In (Eqn.~\ref{eq:Ldeconf2d}) {\it Sign}= $\pm$ is introduced which reflects the sign
ambiguity in the partition functional approach.
%A check has been done whether the choice of {\it Sign} has an impact on the existence
%of the partition functional $Z_G$!
}

The Lagrangian (\ref{eq:Ldeconf2d}) contains up to second-order in $\bar{\lambda}_3$ the interactions 
shown in Fig.~\ref{fig:QandN-graph}, representing the processes
(a) $\lambda_3^0: D  \rightarrow D $, (b) $\lambda_3^2: QN \rightarrow NQ$,
(c) $\lambda_3^1: DQ \rightarrow N$, (d) $\lambda_3^1: N \rightarrow DQ$, where $D= QQ$.

%%--------------------------------------------------------------------------
\noindent The partition functional becomes, see \cite{Kap89},
%{\red check (-)-sign after exponential}
\begin{eqnarray}
	Z_G &=& \int [d\bar{\psi}] [d\psi] [d\bar{q}] [dq] [d\sigma] [d\omega_\mu]
	\int {\cal D}\chi^{a \mu} {\cal D}{\chi}_\mu^{a \dagger}
	\exp \biggl[ \int_0^\beta d\tau \int d^3x \cdot\nonumber\\ && \times
	 \bigl( {\cal L}_N + {\cal L}_Q 
	 + {\cal L}_M + {\cal L}(\chi,q,\psi)
	+\mu_N \psi^\dagger \psi + \mu_Q q^\dagger q \bigr)\biggr],
%	+ \mu_{D} \chi_\mu^{a\dagger}\chi^{a \mu} \bigr)\biggr], 
\label{eq:Mix.63}\end{eqnarray}
where ${\cal L}_\chi = {\cal L}^{(0)}+{\cal L}^{(1)}_{int}$.
%--------------------------------------------------------------------------
 Shifting $\chi^a_\mu \rightarrow A^a_\mu$ and because ${\cal D}\chi^a_\mu = {\cal D}A^a_\mu$ etc
 the exponent is quadratic in $A^a_\mu$, and the $A^a_\mu$ integrals can be performed giving a
 constant factor. Then, ${\cal L}(\chi,q,\psi) \rightarrow {\cal L}^{(2)}_{int}$ in the exponent,
 and the Lagrangian in $Z_G$ becomes ${\cal L} = {\cal L}_N+{\cal L}_Q+{\cal L}_M+{\cal L}^{(2)}_{int}$
 with 
 \begin{eqnarray}
 	{\cal L}^{(2)}_{int} &=& -{\it Sign}\ \bar{\lambda}_3^2\ B_\mu^\dagger B^\mu = 
 -{\it Sign}\ \bar{\lambda}_3^2\ \bigl(\bar{\psi}\gamma_5\gamma_\mu q\bigr) 
 \bigl(\bar{q}\gamma_5\gamma^\mu\psi\bigr),    
 \label{eq:Mix.64}\end{eqnarray}
 which leads to the same contact-interaction potential as in \cite{YYR24}
 \footnote{This is a correction of the Appendix in \cite{YYR24}. We stress that
 the used potential $V^{(qN)}_{DQB}(r)$ is correct as well as the results in \cite{YYR24}.}.

 \noindent {\it The sign ambiguity in the partition functional approach
 is not present in the "diquark-propagator" calculation. Therefore, the proof of the
 strong repulsion from diquark-exchange is given by the "diquark-propagator" derivation.\\
 In order to agree with the interaction in Eqn.~(\ref{eq:Diqrk.2a}) we must choose $Sign=-1$.
 }
%\end{minipage} }\\
%\end{center}
%--------------------------------------------------------------------------
 \begin{acknowledgments}
 I am very grateful to Prof. Yasuo Ymamamoto for our collaboration 
 on the baryon and quark interactions.
 \end{acknowledgments}
%--------------------------------------------------------------------------------------------------
  {99}

 \end{document}